%% file: dr12doubleprobe_v1.3.tex
\documentclass[useAMS,usenatbib]{mn2e}
\usepackage{mn2e-breakabs}
\usepackage{graphicx}
\usepackage{subfigure}
\usepackage{times}
\usepackage{caption}
\usepackage{rotating}
\usepackage{rotfloat}
\usepackage{multicol}
\usepackage{array}
\usepackage{booktabs}
\usepackage{color}
\usepackage{afterpage}
\usepackage[multidot]{grffile}
\usepackage{bm}
\usepackage{float}
\usepackage{morefloats}
\usepackage{hyperref}
\usepackage{amsmath}
\usepackage{mathrsfs}
\providecommand{\adsurl}[1]{\href{#1}{ADS}}

\newcommand{\Hunit}{\,{\rm km}\,{\rm s}^{-1}\,{\rm Mpc}^{-1}}

\def\la{\mathrel{\mathpalette\fun <}}
\def\ga{\mathrel{\mathpalette\fun >}}
\def\fun#1#2{\lower3.6pt\vbox{\baselineskip0pt\lineskip.9pt
        \ialign{$\mathsurround=0pt#1\hfill##\hfil$\crcr#2\crcr\sim\crcr}}}

\newcommand{\be}{\begin{equation}}
\newcommand{\ee}{\end{equation}}
\newcommand{\ba}{\begin{eqnarray}}
\newcommand{\ea}{\end{eqnarray}}
\newcommand{\simgt}{\,\hbox{\lower0.6ex\hbox{$\sim$}\llap{\raise0.6ex\hbox{$>$}}}\,}
\newcommand{\simlt}{\,\hbox{\lower0.6ex\hbox{$\sim$}\llap{\raise0.6ex\hbox{$<$}}}\,}

\begin{document}

\title[Double-Probe Measurements from BOSS \& Planck]
{The clustering of galaxies in the completed SDSS-III Baryon Oscillation Spectroscopic Survey:
double-probe measurements from BOSS galaxy clustering \& Planck data
-- towards an analysis without informative priors
}

\author[Pellejero-Ibanez et al.]{
  \parbox{\textwidth}{
Marcos Pellejero-Ibanez$^{1,2,3,4}$\thanks{E-mail: mpi@iac.es},
 Chia-Hsun Chuang$^{3,4}$\thanks{E-mail: achuang@aip.de},
 J. A. Rubi\~no-Mart\'{\i}n$^{1,2}$,
 Antonio J. Cuesta$^{5}$,
 Yuting Wang$^{6,7}$,
 Gong-bo Zhao$^{6,7}$,
 Ashley J. Ross$^{8,7}$,
Sergio~Rodr\'iguez-Torres$^{3,9,10}$,
Francisco Prada$^{3,9,11}$,
An\v{z}e Slosar$^{12}$, 
Jose A. Vazquez$^{12}$,
Shadab Alam$^{13,14}$,
Florian Beutler$^{15,7}$,
Daniel J. Eisenstein$^{16}$,
H\'ector Gil-Mar\'{i}n$^{17,18}$,
Jan Niklas Grieb$^{19,20}$,
Shirley Ho$^{13,14,15,21}$,
Francisco-Shu Kitaura$^{4,15,21}$,
Will J. Percival$^{7}$,
Graziano Rossi$^{22}$,
Salvador Salazar-Albornoz$^{19,20}$,
Lado Samushia$^{23,24,7}$,
Ariel G. S\'anchez$^{20}$,
Siddharth Satpathy$^{13,14}$,
Hee-Jong Seo$^{25}$,
Jeremy L. Tinker$^{26}$,
Rita Tojeiro$^{27}$,
Mariana Vargas-Maga\~na$^{28}$,
Joel R. Brownstein$^{29}$, 
Robert C Nichol$^{7}$,
Matthew D Olmstead$^{30}$
}
  \vspace*{4pt} \\
$^{1}$ Instituto de Astrof\'isica de Canarias (IAC), C/V\'ia L\'actea, s/n, E-38200, La Laguna, Tenerife, Spain\\
$^{2}$ Departamento Astrof\'isica, Universidad de La Laguna (ULL), E-38206 La Laguna, Tenerife, Spain\\
$^3$ Instituto de F\'{\i}sica Te\'orica, (UAM/CSIC), Universidad Aut\'onoma de Madrid,  Cantoblanco, E-28049 Madrid, Spain \\
$^{4}$ Leibniz-Institut f\"ur Astrophysik Potsdam (AIP), An der Sternwarte 16, 14482 Potsdam, Germany\\
$^{5}$ Institut de Ci{\`e}ncies del Cosmos (ICCUB), Universitat de Barcelona (IEEC-UB), Mart{\'\i} i Franqu{\`e}s 1, E08028 Barcelona, Spain\\
$^{6}$ National Astronomy Observatories, Chinese Academy of Science, Beijing, 100012, P.R.China\\
$^{7}$ Institute of Cosmology and Gravitation, University of Portsmouth, Dennis Sciama Building, Portsmouth PO1 3FX, UK\\ 
$^{8}$ Center for Cosmology and Astroparticle Physics, Department of Physics, The Ohio State University, OH 43210, USA\\ 
$^{9}$ Campus of International Excellence UAM+CSIC, Cantoblanco, E-28049 Madrid, Spain\\ 
$^{10}$ Departamento de F\'isica Te\'orica M8, Universidad Autonoma de Madrid (UAM), Cantoblanco, E-28049, Madrid, Spain\\ 
$^{11}$ Instituto de Astrof\'{\i}sica de Andaluc\'{\i}a (CSIC), Glorieta de la Astronom\'{\i}a, E-18080 Granada, Spain \\ 
$^{12}$ Brookhaven National Laboratory, Upton, NY 11973\\ 
$^{13}$ Department of Physics, Carnegie Mellon University, 5000 Forbes Ave., Pittsburgh, PA 15213, USA\\ 
$^{14}$ The McWilliams Center for Cosmology, Carnegie Mellon University, 5000 Forbes Ave., Pittsburgh, PA 15213\\ 
$^{15}$ Lawrence Berkeley National Lab, 1 Cyclotron Rd, Berkeley CA 94720, USA\\ 
$^{16}$ Harvard-Smithsonian Center for Astrophysics, 60 Garden St., Cambridge, MA 02138, USA\\ 
$^{17}$ Sorbonne Universités, Institut Lagrange de Paris (ILP), 98 bis Boulevard Arago, 75014 Paris, France \\ 
$^{18}$ Laboratoire de Physique Nucléaire et de Hautes Energies, Universit\'e Pierre et Marie Curie, 4 Place Jussieu, 75005 Paris, France \\
$^{19}$ Universit\"ats-Sternwarte M\"unchen, Scheinerstrasse 1, 81679, Munich, Germany\\ 
$^{20}$ Max-Planck-Institut f\"ur extraterrestrische Physik, Postfach 1312, Giessenbachstr., 85741 Garching, Germany\\ 
$^{21}$ Departments of Physics and Astronomy, University of California, Berkeley, CA 94720, USA\\ 
$^{22}$Department of Physics and Astronomy, Sejong University, Seoul, 143-747, Korea\\ 
$^{23}$ Kansas State University, Manhattan KS 66506, USA\\          
$^{24}$ National Abastumani Astrophysical Observatory, Ilia State University, 2A Kazbegi Ave., GE-1060 Tbilisi, Georgia\\  
$^{25}$Department of Physics and Astronomy, Ohio University, 251B Clippinger Labs, Athens, OH 45701, USA\\ 
$^{26}$ Center for Cosmology and Particle Physics, Department of Physics, New York University, 4 Washington Place, New York, NY 10003, USA\\ 
$^{27}$ School of Physics and Astronomy, University of St Andrews, St Andrews, KY16 9SS, UK\\ 
$^{28}$ Instituto de F\'{\i}sica, Universidad Nacional Aut\'onoma de M\'exico, Apdo. Postal 20-364, M\'exico\\ 
$^{29}$ Department of Physics and Astronomy, University of Utah, 115 S 1400 E, Salt Lake City, UT 84112, USA \\ 
$^{30}$ Department of Chemistry and Physics, King's College, 133 North River St, Wilkes Barre, PA 18711, USA\\ 
}

\date{\today} 

\maketitle

\begin{abstract}
We develop a new methodology called double-probe analysis with the aim of minimizing informative priors in the estimation of cosmological parameters. 
Using our new methodology,
we extract the dark-energy-model-independent cosmological constraints from the joint data sets of Baryon Oscillation Spectroscopic 
Survey (BOSS) galaxy sample and Planck cosmic microwave background (CMB) measurement.
We measure the mean values and covariance matrix of 
$\{R$, $l_a$, $\Omega_b h^2$, $n_s$, $log(A_s)$, $\Omega_k$, $H(z)$, $D_A(z)$, $f(z)\sigma_8(z)\}$, which give an efficient summary of 
Planck data and 2-point statistics from BOSS galaxy sample. The CMB shift parameters are 
$R=\sqrt{\Omega_m H_0^2}\,r(z_*)$, and $l_a=\pi r(z_*)/r_s(z_*)$, where $z_*$ is the redshift at the last scattering surface, and 
$r(z_*)$ and $r_s(z_*)$ denote our comoving distance to $z_*$ and sound horizon at $z_*$ respectively; $\Omega_b$ is the baryon fraction 
at $z=0$.
The advantage of this method is that we do not need to 
put informative priors on the cosmological parameters that galaxy clustering is not able to constrain well, i.e. $\Omega_b h^2$ and $n_s$. 

Using our double-probe results, we obtain 
$\Omega_m=0.304\pm0.009$, $H_0=68.2\pm0.7$, and $\sigma_8=0.806\pm0.014$ assuming $\Lambda$CDM; 
$\Omega_k=0.002\pm0.003$ assuming oCDM; $w=-1.04\pm0.06$ assuming $w$CDM; 
$\Omega_k=0.002\pm0.003$ and $w=-1.00\pm0.07$ assuming o$w$CDM;
and $w_0=-0.84\pm0.22$ and $w_a=-0.66\pm0.68$ assuming $w_0w_a$CDM.
The results show no tension with the flat $\Lambda$CDM cosmological paradigm.
By comparing with the full-likelihood analyses 
with fixed
dark energy models, we demonstrate that the double-probe method provides robust 
cosmological parameter constraints which can be conveniently used to study dark energy models.

We extend our study to measure the sum of neutrino mass using different methodologies including double probe analysis (introduced in 
this study), the full-likelihood analysis, and single probe analysis. From the double probe analysis, we obtain
$\Sigma m_\nu<0.10/0.22$ (68\%/95\%) assuming $\Lambda$CDM and
$\Sigma m_\nu<0.26/0.52$ (68\%/95\%) assuming $w$CDM.
This paper is part of a set that analyses the final galaxy clustering
dataset from BOSS.
\end{abstract}


\section{Introduction} \label{sec:intro}

We have entered the era of precision cosmology along with the dramatically increasing amount of sky surveys, including the cosmic microwave 
background (CMB; e.g., \citealt{Bennett:2012zja,Ade:2013sjv}), supernovae (SNe; \citealt{Riess:1998cb,Perlmutter:1998np}), weak lensing (e.g., see \citealt{VanWaerbeke:2003uq} for a review), and large-scale structure from galaxy redshift surveys, e.g. 2dF Galaxy Redshift Survey (2dFGRS; \citealt{Colless:2001gk,Colless:2003wz}, Sloan Digital Sky Survey (SDSS, \citealt{York:2000gk,Abazajian:2008wr}, WiggleZ \citep{Drinkwater:2009sd, Parkinson:2012vd}, and the Baryon Oscillation Spectroscopic Survey (BOSS; 
\citealt{Dawson:2012va,Alam:2015mbd}) of the SDSS-III \citep{Eisenstein:2011sa}. 
The future galaxy redshift surveys, e.g. Euclid\footnote{http://sci.esa.int/euclid} \citep{Laureijs:2011gra}, Dark Energy Spectroscopic Instrument \footnote{http://desi.lbl.gov/} (DESI; \citealt{Schlegel:2011zz}), and WFIRST\footnote{http://wfirst.gsfc.nasa.gov/} \citep{Green:2012mj}, will collect data at least an order of magnitude more. It is critical to develop the methodologies which could reliably extract the cosmological information from such large amount of data. 

The galaxy redshifts samples have been analysed studied in a cosmological context (see, e.g., 
\citealt{Tegmark:2003uf,Hutsi:2005qv,Padmanabhan:2006cia,Blake:2006kv,Percival:2007yw,Percival:2009xn,Reid:2009xm,
Montesano:2011bp, Eisenstein:2005su,Okumura:2007br,Cabre:2008sz,Martinez:2008iu,Sanchez:2009jq,Kazin:2009cj,Chuang:2010dv,Samushia:2011cs,Padmanabhan:2012hf,Xu:2012fw,
Anderson:2012sa,Manera:2012sc,Nuza:2012mw,Reid:2012sw,Samushia:2012iq,Tojeiro:2012rp, Anderson:2013oza, Chuang:2013hya, 
Sanchez:2013uxa, Kazin:2013rxa,Wang:2014qoa,Anderson:2013zyy,Beutler:2013yhm,Samushia:2013yga,Chuang:2013wga,Sanchez:2013tga,
Ross:2013vla,Tojeiro:2014eea,Reid:2014iaa,Alam:2015qta,Gil-Marin:2015nqa,Gil-Marin:2015sqa,Cuesta:2015mqa}).

\cite{Eisenstein:2005su} demonstrated the feasibility of measuring $\Omega_mh^2$ and an effective distance, $D_V(z)$ 
from the SDSS DR3 \citep{Abazajian:2004it} LRGs, where $D_V(z)$ corresponds to a combination of Hubble expansion rate $H(z)$ and angular-diameter distance $D_A(z)$.
\cite{Chuang:2011fy} demonstrated the feasibility of measuring $H(z)$ and $D_A(z)$ simultaneously using the galaxy clustering data from the two dimensional two-point correlation function of SDSS DR7 \citep{Abazajian:2008wr} LRGs and it has been improved later on in
\cite{Chuang:2012ad,Chuang:2012qt} upgrading the methodology and modelling to measure $H(z)$, $D_A(z)$, the normalized growth rate $f(z)\sigma_8(z)$, and the physical matter density
$\Omega_m h^2$ from the same data.
Analyses have been perform to measure $H(z)$, $D_A(z)$, and $f(z)\sigma_8(z)$ from earlier data release of SDSS BOSS galaxy sample 
\cite{Reid:2012sw,Chuang:2013hya,Wang:2014qoa,Anderson:2013zyy,Beutler:2013yhm,Chuang:2013wga,Samushia:2013yga}.

There are some cosmological parameters, e.g. $\Omega_bh^2$ (the physical baryon fraction) and $n_s$ (the scalar index of the power law primordial fluctuation), not well constrained by galaxy clustering analysis. We usually use 
priors adopted from CMB measurements or fix those to the best fit values obtained from CMB while doing Markov Chain Monte Carlo (MCMC) analysis. 
There would be some concern of missing weak degeneracies between these parameters and those measured. These could lead to incorrect constraints if models
with very different predictions are tested, or double-counting when
combining with CMB measurements.
One might add some 
systematics error budget to be safe from the potential bias (e.g., see \cite{Anderson:2013zyy}). An alternative approach is to use a very wide 
priors, e.g. 5 or 10 $\sigma$ flat priors from CMB, to minimize the potential systematics bias from priors (e.g., see \cite{Chuang:2010dv,Chuang:2011fy}). However, the 
approach would obtain weaker constraints due to the wide priors.
In this study, we test the ways in which LSS constraints are combined with CMB data,
focussing on the information content, and the priors used when analysing
LSS data. Since CMB data can be summarized with few parameters 
(e.g., see \cite{Wang:2007mza}), we use the joint data set from Planck and BOSS to extract the cosmological constraints without 
fixing dark energy models. By combining the CMB data and the
BOSS data in the upstream of the data analysis to constrain the
cosmological constraints, we call our method "double-probe analysis". Our
companion paper,  \cite{Chuang16}, constrains geometric and growth information
from the BOSS data alone independent of the CMB data, thereby dubbed
"single-probe", and combines with the CMB data in the downstream of the
analysis.
Note that we assume there is no early time dark energy or dark energy clustering in this study. $\Omega_bh^2$ and $n_s$ will be well 
constrained by CMB so that we will obtain the cosmological constraints without concerning the problem of priors. The only input 
parameter which is not well constrained by our analysis is the galaxy bias on which is applied a wide flat prior. In principle, our 
methodology extract the cosmological constraints from the joint data set with the optimal way since we do not need to include the 
uncertainty introduced by the priors.

In addition to constraining dark energy model parameters, we extend our study to constrain neutrino masses.
High energy physics experiments provides with the squared of mass differences between neutrino species from oscillation neutrino 
experiments. Latest results are $\Delta m^2_{21}=7.53\pm 0.18 \times 10^{-5} eV^2$ and $\Delta m^2_{32}=2.44\pm 0.06 \times 10^{-3} eV^2$
for the normal hierarchy ($m_3\gg m_2 \simeq m_1$) and $\Delta m^2_{32}=2.52\pm 0.07 \times 10^{-3} eV^2$ for the inverted mass hierarchy
($m_3\ll m_2 \simeq m_1$) (\citealt{1674-1137-38-9-090001}).
Cosmology shows as a unique tool for the measurement of the sum of neutrino masses $\Sigma m_{\nu}$, since this quantity affects the 
expansion rate and the way structures form and evolve.
$\Sigma m_\nu$ estimations from galaxy clustering has been widely studied theoretically (see \citealt{Hu:1997mj,Lesgourgues:2006nd} for a review)
and with different samples such as WiggleZ (see \citealt{Riemer-Sorensen:2013jsa,Cuesta:2015iho}) or SDSS data
(see \citealt{Aubourg:2014yra,Beutler:2014yhv,Reid:2009nq,Thomas:2009ae,Zhao:2012xw}).
At late times, massive neutrinos can damp
the formation of cosmic structure on small scales due to the
free-streaming effect \citep{Dolgov:2002wy}.
Existing in the form of radiation in the early Universe, neutrinos shift the
epoch of the matter-radiation equality thus changing the shape of the cosmic microwave background (CMB) angular
power spectrum.
They affect CMB via 
the so called Early Integrated Sachs Wolfe Effect and they influence gravitational lensing measurements (e.g., see \citealt{Lesgourgues:2005yv}).
Recent publications have attempted to constrain $\Sigma m_\nu$ , imposing
upper limits 
\citep{Seljak:2006bg,Hinshaw:2008kr,Dunkley:2008ie,Reid:2009nq,Komatsu:2010fb,Saito3,Tereno:2008mm,Gong:2008pg,Ichiki:2008ye,Li:2008vf,dePutter:2012sh,Xia:2012na,Sanchez:2012sg,Giusarma:2013pmn}
and some hints of lower limits using cluster abundance results 
\citep{Ade:2013lmv,Battye:2013xqa,Wyman:2013lza,Burenin:2013wg,Rozo:2013hha}.
We measure the sum of neutrino mass using different methodologies including double probe analysis (introduced in 
this study), the full-likelihood analysis, and single probe analysis (\citealt{Chuang16}; companion paper).

This paper is organized as follows. 
In Section \ref{sec:data}, we introduce the Planck data, the SDSS-III/BOSS DR12 galaxy sample and mock catalogues used 
in our study. 
In Section \ref{sec:method}, we describe the details of the 
methodology that constrains cosmological parameters from our joint CMB and galaxy clustering analysis. 
In Section \ref{sec:results}, we present our double-probe cosmological measurements. 
In Section \ref{sec:use}, we demonstrate how to derive cosmological constraints from our measurements with some given dark energy model.
In Section \ref{sec:full_run}, opposite to the manner of dark energy model independent method, we present the results from the full-likelihood analysis  with fixing dark energy models.
In Section \ref{sec:mnu}, we measure the sum of neutrino mass with different methodologies.
We summarize and conclude in Section \ref{sec:conclusion}.

\section{Data sets \& mocks} 
\label{sec:data}

\subsection{The SDSS-III/BOSS Galaxy Catalogues}
\label{sec:galaxy}
The Sloan Digital Sky Survey (SDSS; \citealt{Fukugita:1996qt,Gunn:1998vh,York:2000gk,Smee:2012wd}) mapped over one quarter 
of the sky using the dedicated 2.5 m Sloan Telescope \citep{Gunn:2006tw}.
The Baryon Oscillation Sky Survey (BOSS, \citealt{Eisenstein:2011sa, Bolton:2012hz, Dawson:2012va}) is part of the SDSS-III survey. 
It is collecting the spectra and redshifts for 1.5 million galaxies, 160,000 quasars and 
100,000 ancillary targets. The Data Release 12 \citep{Alam:2015mbd} has been made publicly available\footnote{http://www.sdss3.org/}.
We use galaxies from the SDSS-III BOSS DR12 CMASS catalogue in the redshift range $0.43<z<0.75$ and LOWZ catalogue in the range $0.15<z<0.43$.
CMASS samples are selected with an approximately constant stellar mass threshold \citep{Eisenstein:2011sa}; 
LOWZ sample consists of red galaxies at $z<0.4$ from the SDSS DR8 \citep{Aihara:2011sj} image data.
We are using 800853 CMASS galaxies and  361775 LOWZ galaxies.
The effective redshifts of the sample are $z=0.59$ and $z=0.32$ respectively.
The details of generating this sample are described in \cite{Reid:2015gra}.

\subsection{The Planck Data}
\textit{Planck} \citep{tauber2010,planck2011-1} is the third
generation space mission, following COBE and WMAP, to
measure the anisotropy of the CMB. It observed the sky in nine frequency bands
covering the range 30--857\,GHz with high sensitivity and angular resolutions from 31' to
5'.  The Low Frequency Instrument \citep[LFI;][]{Bersanelli:2010qb,Mennella:2011ay}
covers the bands centred at 30, 44, and 70\,GHz using pseudo-correlation radiometers detectors, while
the High Frequency Instrument \citep[HFI;][]{planck2011-1.5} covers the 100, 143, 217, 353, 545,
and 857\,GHz bands with bolometers. Polarisation is measured in all
but the highest two bands \citep{leahy2010,rosset2010}.
In this paper, we used the 2015 \textit{Planck} release
\citep{planck2015-i}, which included the
full mission maps and associated data products.

\subsection{The Mock Galaxy Catalogues}
We use 2000 BOSS DR12 MultiDark-PATCHY (MD-PATCHY) mock galaxy catalogues \citep{Kitaura:2015uqa} for validating our methodology and estimating the covariance matrix in this study. These mock catalogues were constructed using a similar procedure described in \citealt{Rodriguez-Torres:2015vqa} where they constructed the BOSS DR12 lightcone mock catalogues using the MultiDark $N$-body simulations. However, instead of using $N$-body simulations, the 2000 MD-PATCHY mocks catalogues were constructed using the PATCHY approximate simulations.
These mocks are produced using ten boxes at different redshifts that are created with the PATCHY-code \citep{Kitaura:2013cwa}. The PATCHY-code is composed of two parts: 1) computing approximate dark matter density field; and 2) populating galaxies from dark matter density field with the biasing model. The dark matter density field is estimated using Augmented Lagrangian Perturbation Theory (ALPT; \cite{Kitaura:2012tj}) which combines the second order perturbation theory (2LPT; e.g. see \cite{Buchert:1993ud,Bouchet:1994xp,Catelan:1994ze}) and spherical collapse approximation (see \cite{Bernardeau:1993ac,Mohayaee:2005xm,Neyrinck:2012bf}). The biasing model includes deterministic bias and stochastic bias (see \cite{Kitaura:2013cwa,Kitaura:2014mja} for details). The velocity field is constructed based on the displacement field of dark matter particles. The modeling of finger-of-god has also been taken into account statistically. The mocks match the clustering of the galaxy catalogues for each redshift bin (see \cite{Kitaura:2015uqa} for details) and have been used in recent galaxy clustering studies \citep{Cuesta:2015mqa,Gil-Marin:2015nqa,Gil-Marin:2015sqa,Rodriguez-Torres:2015vqa,Slepian:2015hca} and void clustering studies \citep{Kitaura:2015ubm,Liang:2015oqc}.  
They are also used in 
\cite{Acacia} (BOSS collaboration paper for final data release) and its companion papers (this paper and 
\cite{Ross16, 
Vargas-Magana16, 
Beutler16b, 
Satpathy16, 
Beutler16c, 
Sanchez16a, 
Grieb16, 
Sanchez16b, 
Chuang16, 
Slepian16a, 
Slepian16b, 
Salazar-Albornoz16, 
Zhao16, 
Wang16}

\section{Methodology} 
\label{sec:method}
We develop a new methodology to extract the cosmological constraints from the joint data set of the Planck CMB data and BOSS galaxy clustering measurements fitting the LSS data with parameter combinations defining the key cosmological dependencies, while including CMB constraints to simultaneously constrain other parameters.
This means that we can define constraints that can subsequently be used to
constrain a wide-range of Dark Energy models. Similar approaches have been applied to these data separately. Our work is
the first to investigate how in detail this joint analysis should be
performed.

\subsection{Likelihood from BOSS galaxy clustering}
In this section, we describe the steps to compute the likelihood from the BOSS galaxy clustering.

\subsubsection{Measure Multipoles of the Two-Point Correlation Function}  \label{sec:multipoles}
We convert the measured redshifts of the BOSS CMASS and LOWZ galaxies to comoving distances 
by assuming a fiducial model, i.e., flat $\Lambda$CDM with $\Omega_m=0.307115$ and $h=0.6777$ 
which is the same model adopted for constructing the mock catalogues (see \cite{Kitaura:2015uqa}
). 
To compute the two-dimensional two-point correlation function, we use the two-point correlation function estimator given by 
\cite{Landy:1993yu}:
\begin{equation}
\label{eq:xi_Landy}
\xi(s,\mu) = \frac{DD(s,\mu)-2DR(s,\mu)+RR(s,\mu)}{RR(s,\mu)},
\end{equation}
where $s$ is the separation of a pair of objects and $\mu$ is the cosine of the angle between the directions between the line of sight (LOS) and the line connecting the pair the objects. DD, DR, and RR represent the normalized data-data,
data-random, and random-random pair counts, respectively, for a given
distance range. The LOS is defined as the direction from the observer to the 
centre of a galaxy pair. Our bin size is
$\Delta s=1 \, h^{-1}$Mpc and $\Delta \mu=0.01$. 
The Landy and Szalay estimator has minimal variance for a Poisson
process. Random data are generated with the same radial
and angular selection functions as the real data. One can reduce the shot noise due
to random data by increasing the amount of random data. The number
of random data we use is about 50 times that of the real data. While
calculating the pair counts, we assign to each data point a radial
weight of $1/[1+n(z)\cdot P_w]$, where $n(z)$ is the radial
number density and $P_w = 1\cdot 10^4$ $h^{-3}$Mpc$^3$ (see  
\citealt{Feldman:1993ky}).

The traditional multipoles of the two-point correlation function, in redshift space, are defined by
\ba
\label{eq:multipole_1}
\xi_l(s) &\equiv & \frac{2l+1}{2}\int_{-1}^{1}{\rm d}\mu\, \xi(s,\mu) P_l(\mu)\nonumber,
\ea
where $P_l(\mu)$ is the Legendre Polynomial ($l=$0 and 2 here). 
We integrate over a spherical shell with radius $s$,
while actual measurements of $\xi(s,\mu)$ are done in discrete bins.
To compare the measured $\xi(s,\mu)$ and our theoretical model, the last integral in Eq.(\ref{eq:multipole_1}) should be converted into a sum,
\begin{equation}\label{eq:multipole}
 \hat{\xi}_l(s) \equiv \frac{\displaystyle\sum_{s-\frac{\Delta s}{2} < s' < s+\frac{\Delta s}{2}}\displaystyle\sum_{0\leq\mu\leq1}(2l+1)\xi(s',\mu)P_l(\mu)}{\mbox{Number of bins used in the numerator}},
\end{equation}
where $\Delta s=5$ $h^{-1}$Mpc in this work.

\begin{figure*}
\centering
\subfigure{\includegraphics[width=1 \columnwidth,clip,angle=-0]{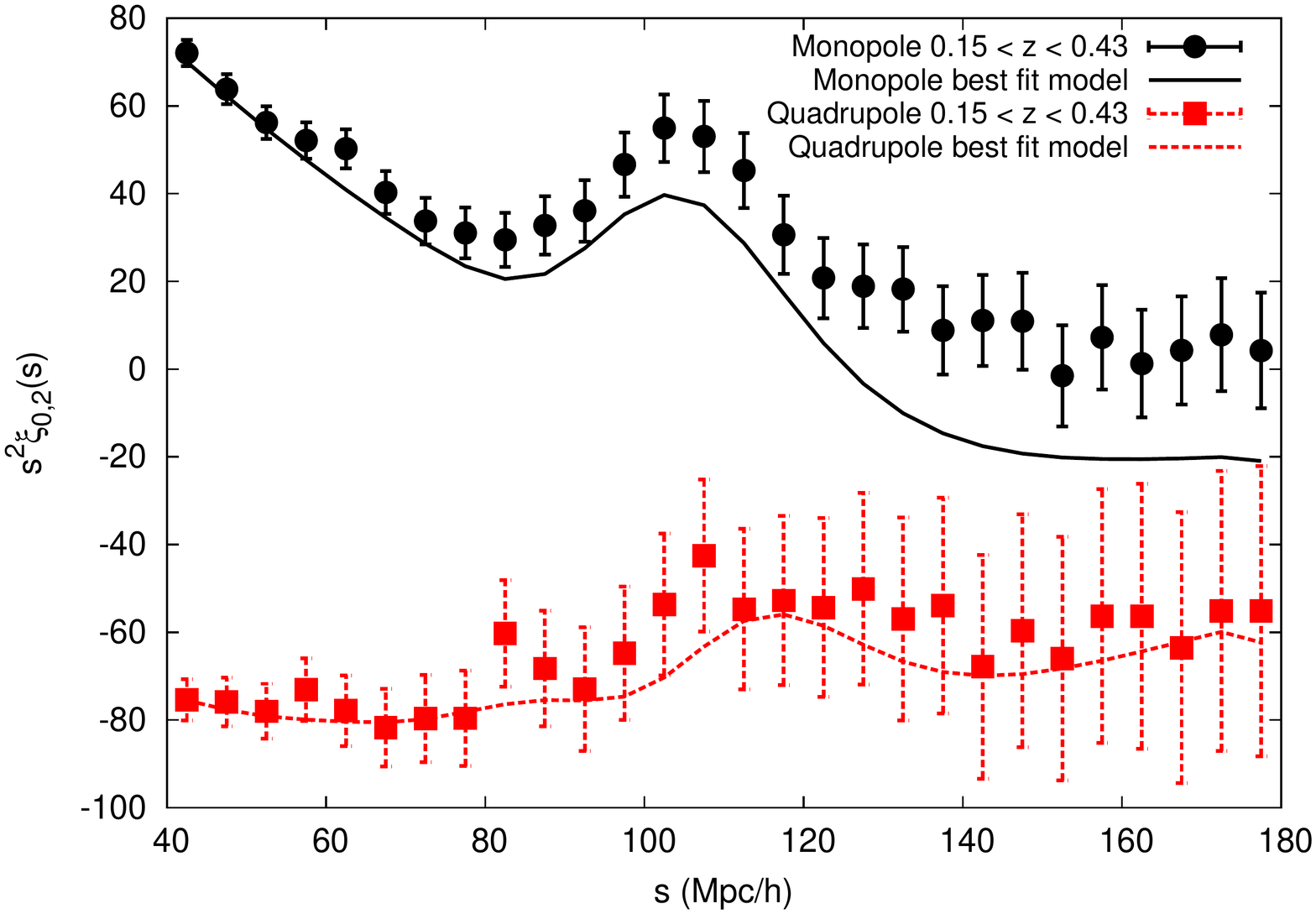}}
\subfigure{\includegraphics[width=1 \columnwidth,clip,angle=-0]{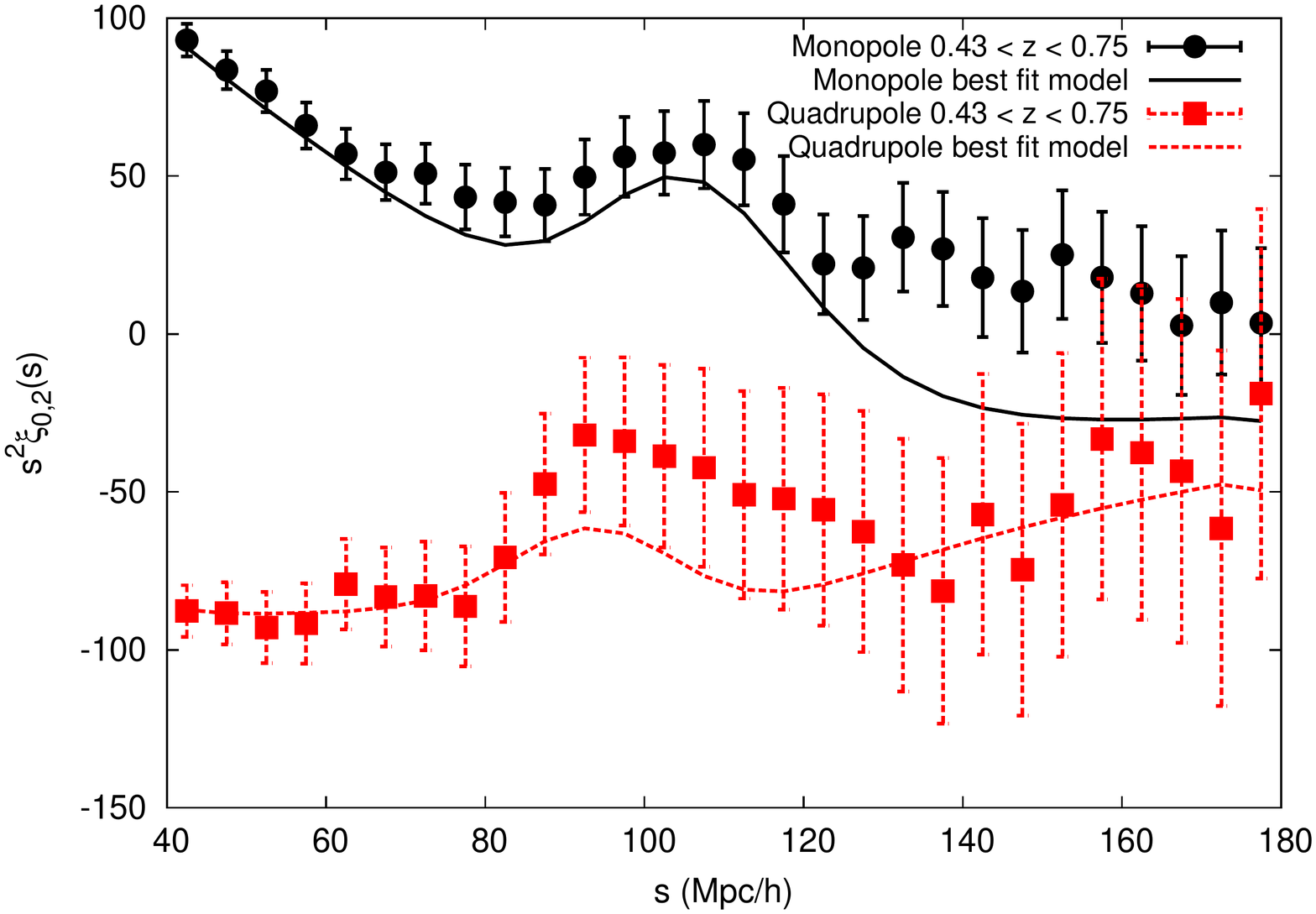}}
\caption{
Left panel: measurement of monopole and quadrupole of the correlation function from the BOSS DR12 LOWZ galaxy sample within $0.15<z<0.43$ compared to the best fited theoretical models  (solid lines).
Right panel: measurement of effective monopole and quadrupole of the correlation function from the BOSS DR12 CMASS galaxy sample within $0.43<z<0.75$ compared to the best fitted theoretical models  (solid lines).
The error bars are the square roots of the diagonal elements of the covariance matrix.
In this study, our fitting scale ranges are $40h^{-1}$Mpc $<s<180h^{-1}$Mpc; the bin size is $5h^{-1}$Mpc.
}
\label{fig:mp_cmass_lowz}
\end{figure*}

Fig.\ref{fig:mp_cmass_lowz} shows the monopole ($\hat{\xi}_0$) and quadrupole ($\hat{\xi}_2$) measured from the BOSS CMASS and LOWZ galaxy sample compared with the best fit theoretical models.

We are using the scale range $s=40-180\,h^{-1}$Mpc and the bin size is 5 $h^{-1}$Mpc. 
The data points from the multipoles in the scale range considered are combined to form a 
vector, $X$, i.e.,
\be
{\bf X}=\{\hat{\xi}_0^{(1)}, \hat{\xi}_0^{(2)}, ..., \hat{\xi}_0^{(N)}; 
\hat{\xi}_2^{(1)}, \hat{\xi}_2^{(2)}, ..., \hat{\xi}_2^{(N)};...\},
\label{eq:X}
\ee
where $N$ is the number of data points in each measured multipole; here $N=28$.
The length of the data vector ${\bf X}$ depends on the number of multipoles used. 

\subsubsection{Theoretical Two-Point Correlation Function}
\label{sec:model_2PCF}
Following \cite{Chuang16}, companion paper, we use two models to compute the likelihood of the galaxy clustering measurements. One is a 
fast model which is used to narrow down the parameters space scanned; the other is a slower model which is used to calibrate the results 
from the fast model.\\

Fast model: 
The fast model we use is the two-dimensional dewiggle model explained in \cite{Chuang16}, companion paper.
The theoretical model can be constructed by first and higher order perturbation theory. 
We first adopt the cold dark matter model and the simplest inflation model (adiabatic initial condition).
Computing the linear matter power spectra, $P_{lin}(k)$, by using CAMB (Code for Anisotropies in the Microwave Background,
\citealt{Lewis:1999bs}) we can decomposed it into two parts:
\begin{equation} \label{eq:pk_lin}
P_{lin}(k)=P_{nw}(k)+P_{BAO}^{lin}(k),
\end{equation}
where $P_{nw}(k)$ is the ``no-wiggle'' power spectrum calculated using Eq.(29) from \cite{Eisenstein:1997ik} and $P_{BAO}^{lin}(k)$ is the 
``wiggled'' part defined by previous Eq. (\ref{eq:pk_lin}).
Nonlinear damping effect of the ``wiggled'' part, in redshift space, is approximated following \cite{Eisenstein:2006nj} by
\begin{equation} \label{eq:bao}
P_{BAO}^{nl}(k,\mu_k)=P_{BAO}^{lin}(k)\cdot \exp\left(-\frac{k^2}{2k_\star^2}[1+\mu_k^2(2f+f^2)]\right),
\end{equation}
where $\mu_k$ is the cosine of the angle between ${\bf k}$ and the LOS, $f$ is the growth rate, and
$k_\star$ is computed following \cite{Crocce:2005xz} and \cite{Matsubara:2007wj} by
\begin{equation} \label{eq:kstar}
k_\star=\left[\frac{1}{3\pi^2}\int P_{lin}(k)dk\right]^{-1/2}.
\end{equation}
Thus dewiggled power spectrum is
\begin{equation} \label{eq:pk_dw}
P_{dw}(k,\mu_k)=P_{nw}(k)+P_{BAO}^{nl}(k,\mu_k).
\end{equation}

We include the linear redshift distortion as follows (reference \citep{Kaiser:1987qv}),
\begin{eqnarray} \label{eq:pk_2d}
P_g^s(k,\mu_k)&=&b^2(1+\beta\mu_k^2)^2P_{dw}(k,\mu_k),
\end{eqnarray}
where $b$ is the linear galaxy bias and $\beta$ is the linear redshift distortion parameter.

To compute the theoretical two-point correlation 
function, $\xi(s,\mu)$, we Fourier transform the non-linear power spectrum
$P_g^s(k,\mu_k)$ by using Legendre polynomial expansions and one-dimensional integral convolutions 
as introduced in \cite{Chuang:2012qt}.

We times calibration functions to the fast model by
\begin{eqnarray}
\xi_0^{cal}(s)=(1-e^{-\frac{s}{s_{1}}}+e^{-\left(\frac{s}{s_{2}}\right)^2})\xi_0(s),\\
\xi_2^{cal}(s)=(1-e^{-\frac{s}{s_{3}}}+e^{-\left(\frac{s}{s_{4}}\right)^2})\xi_2(s),
\end{eqnarray}
so that it mimics the slow model presented bellow. We find the calibration parameters, $s_1=12$, $s_2=14$, $s_3=20$, and $s_4=27$, by 
comparing the fast and slow models by visual inspection. 
It is not critical to find the best form of calibration function and its parameters as the model will be callibrated later when 
performing importance sampling with slow model.\\

Slow model: The slower but accurate model we use is "Gaussian streaming model" described in
\cite{Reid:2011ar}. 
The model assumes that the pairwise velocity probability distribution function is
 Gaussian and can be used to relate real space clustering and pairwise
velocity statistics of halos to their clustering in redshift space by  
\begin{eqnarray}
  1+\xi^{s}_{\rm g}(r_{\sigma},r_{\pi}) = \nonumber
\end{eqnarray}
\begin{equation}
\int \left[1+\xi^{r}_{\rm g}(r)\right]
  e^{-[r_\pi - y - \mu v_{12}(r)]^2/2\sigma_{12}^2(r,\mu)} \frac{dy}{\sqrt{2\pi\sigma^2_{12}(r,\mu)}} \label{eq:streamingeqn},
\end{equation}
\noindent
where $r_\sigma$ and $r_\pi$ are the redshift space transverse and LOS distances
between two objects with respect to the observer, $y$ is the {\em real} space
LOS pair separation, $\mu = y/r$, $\xi_{\rm g}^{\rm r}$ is the
real space galaxy correlation function, $v_{12}(r)$ is the average infall velocity of
galaxies separated by real-space distance $r$, and $\sigma_{12}^2(r,\mu)$ is the
rms dispersion of the pairwise velocity between two galaxies separated with
transverse (LOS) real space separation $r_{\sigma}$ ($y$).  
$\xi_{\rm g}^{\rm r}(r)$, $v_{12}(r)$ and $\sigma_{12}^2(r,\mu)$ are computed
in the framework of Lagrangian ($\xi^{\rm r}$) and standard perturbation
theories ($v_{12}$, $\sigma_{12}^2$).  

For large scales, only one nuisance parameter is necessary to describe the clustering of a
sample of halos or galaxies in this model: $b_{1L} = b-1$, the first-order
Lagrangian host halo bias in {\em real} space.
In this study, we consider relative large scales (i.e. $40 < s < 180 h^{-1}$Mpc), so that we do not include 
$\sigma^2_{\rm FoG}$, to model a
velocity dispersion accounting for small-scale motions
of halos and galaxies.
Further details of the model, its numerical implementation, and its accuracy
can be found in \cite{Reid:2011ar}.

\subsubsection{Covariance Matrix} \label{sec:covar}

We use the 2000 mock catalogues created by
\citealt{Kitaura:2015uqa}
for the BOSS DR12 CMASS and LOWZ galaxy sample
to estimate the covariance matrix of the observed correlation function. 
We calculate the multipoles of the correlation functions 
of the mock catalogues and construct the covariance matrix as
\begin{equation}
 C_{ij}=\frac{1}{(N-1)(1-D)}\sum^N_{k=1}(\bar{X}_i-X_i^k)(\bar{X}_j-X_j^k),
\label{eq:covmat}
\end{equation}
where
\begin{equation}
 D=\frac{N_b +1}{N-1},
\label{eq:D}
\end{equation}
$N$ is the number of the mock catalogues, $N_b$ is the number of data bins, $\bar{X}_m$ is the
mean of the $m^{th}$ element of the vector from the mock catalogue multipoles, and
$X_m^k$ is the value in the $m^{th}$ elements of the vector from the $k^{th}$ mock
catalogue multipoles. The data vector ${\bf X}$ is defined by Eq.(\ref{eq:X}).
We also include the correction, $D$, introduced by \cite{Hartlap:2006kj}. 

\subsubsection{Compute Likelihood from Galaxy Clustering}
The likelihood is taken to be proportional to $\exp(-\chi^2/2)$ \citep{press92}, 
with $\chi^2$ given by
\begin{equation} \label{eq:chi2}
 \chi^2\equiv\sum_{i,j=1}^{N_{X}}\left[X_{th,i}-X_{obs,i}\right]
 C_{ij}^{-1}
 \left[X_{th,j}-X_{obs,j}\right]
\end{equation}
where $N_{X}$ is the length of the vector used, 
$X_{th}$ is the vector from the theoretical model, and $X_{obs}$ 
is the vector from the observed data.

As explained in \cite{Chuang:2011fy}, instead of recalculating the observed correlation function while 
computing for different models, we rescale the theoretical correlation function to avoid rendering the $\chi^2$ values arbitrary.
This approach can be considered as an application of Alcock-Paczynski effect \citep{Alcock:1979mp}.
The rescaled theoretical correlation function is computed by
\begin{equation} \label{eq:inverse_theory_2d}
 T^{-1}(\xi_{th}(\sigma,\pi))=\xi_{th}
 \left(\frac{D_A(z)}{D_A^{fid}(z)}\sigma,
 \frac{H^{fid}(z)}{H(z)}\pi\right),
\end{equation}
where $\xi_{th}$ is the theoretical model computed in Sec. \ref{sec:model_2PCF}. Here, $D_A(z)$ and $H(z)$ would be the input parameters and $D_A^{fid}(z)$ and $H^{fid}(z)$ are $\{990.20$Mpc, $80.16\Hunit\}$ at $z=0.32$ (LOWZ) and $\{1409.26$Mpc, $94.09\Hunit\}$ at $z=0.59$ (CMASS).
Then, $\chi^2$ can be rewritten as
\ba 
\label{eq:chi2_2}
\chi^2 &\equiv&\sum_{i,j=1}^{N_{X}}
 \left\{T^{-1}X_{th,i}-X^{fid}_{obs,i}\right\}
 C_{fid,ij}^{-1} \cdot \nonumber\\
 & & \cdot \left\{T^{-1}X_{th,j}-X_{obs,j}^{fid}\right\};
\ea
where $T^{-1}X_{th}$ is the vector computed by eq.\ (\ref{eq:multipole}) from the rescaled theoretical correlation function, eq. (\ref{eq:inverse_theory_2d}).
$X^{fid}_{obs}$ is the vector from observed data measured with the fiducial model (see \citealt{Chuang:2011fy} for more details regarding the rescaling method).

\subsection{Likelihood from Planck CMB data}
Our CMB data set consists of the \textit{Planck} 2015 measurements \citep{planck2015-i, planck2015-xiii}.
The reference likelihood code \citep{planck2015-xi} was downloaded from the
Planck Legacy Archive\footnote{PLA: \url{http://pla.esac.esa.int/}}.
Here we combine the \textit{Plik} baseline likelihood for high multipoles ($30 \le \ell \le 2500$)
using the TT, TE and EE power spectra, and the \textit{Planck} low-$\ell$ multipole likelihood in the
range $2 \le \ell \le 29$ (hereafter lowTEB). We also include the new \textit{Planck} 2015 lensing
likelihood \citep{planck2015-xv}, constructed from the measurements of the power spectrum of the
lensing potential (hereafter referred as "lensing"). We using the \textit{Planck} lensing likelihood, the
$A_{\rm lens}$ parameter is always set to 1 \citep{planck2015-xiii}.

\begin{figure}
\centering
\includegraphics[width=1 \columnwidth,clip,angle=-0]{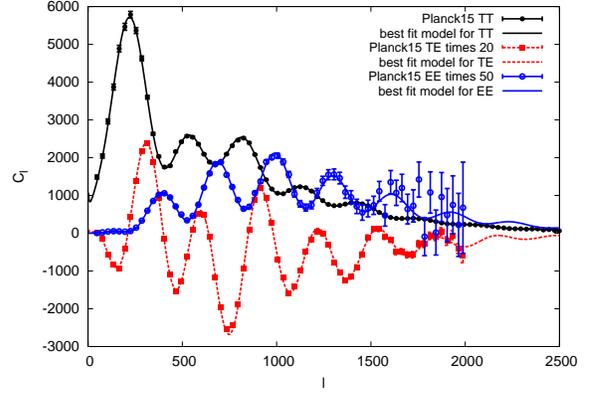}
\caption{Angular power spectrum of temperature and polarization measurement from Planck data and their best fits from our double probe analysis.
}
\label{fig:planck}
\end{figure}

\subsection{Markov Chain Monte-Carlo Likelihood Analysis} 
\label{sec:mcmc}

\subsubsection{basic procedure}
We perform Markov Chain Monte Carlo (MCMC)
likelihood analyses using CosmoMC \citep{Lewis:2002ah,Lewis:2013hha}. 
The fiducial parameter space that we explore spans the parameter set of
$\{\Omega_c h^2$, $\Omega_b h^2$,  $n_s$, $log(A_s)$, $\theta$, $\tau$, $\Omega_k$, $w$,$H(z)$, $D_A(z)$, $\beta(z)$, $b\sigma_8(z)$, $b(z)\}$.
The quantities $\Omega_c$ and $\Omega_b$ are the cold dark matter and
baryon density fractions, $n_s$ is the power-law index of the primordial matter power spectrum, $\Omega_k$ is the curvature density fraction, $w$ is the equation state of dark energy,
$h$ is the dimensionless Hubble
constant ($H_0=100h$ km s$^{-1}$Mpc$^{-1}$), and $\sigma_8(z)$ is the normalization of the power spectrum. Note that, with the joint data set (Planck + BOSS), the only parameter which is not well constrained is $b(z)$. 
We apply a flat prior of $(1,3)$ on it.
The linear redshift distortion parameter can be expressed as $\beta(z)=f(z)/b$.
Thus, one can derive $f(z)\sigma_8(z)$ from the measured $\beta(z)$ and $b\sigma_8(z)$.

\subsubsection{Generate Markov chains with fast model}
We first use the fast model (2D dewiggle model) to compute the likelihood, $\mathcal{L}_{fast}$ and generate the Markov chains. 
The Monte Carlo analysis will go through many random steps keeping or throwing the computed points parameter space according to the Markov likelihood algorithm. Eventually, it will provide the chains of parameter points
with high likelihood describing the constraints to our model.

\subsubsection{Calibrate the likelihood using slow model}
Once we have the fast model
generated chains, we modify the weight of each point by
\begin{equation}
\mathcal{W}_{new}=\mathcal{W}_{old}\frac{\mathcal{L}_{slow}}{\mathcal{L}_{fast}},
\end{equation}
where $\mathcal{L}_{slow}$ and $\mathcal{L}_{fast}$ are the likelihood for given point of input parameters in the chains. We save time by computing only the "important"
points without computing the likelihood of the ones which will not
be included in the first place. The methodology is know as "Importance sampling". However, the typical Importance sampling method is to add likelihood of some additional data set to the given chains, but in this study, we replace the likelihood of a data set.
\\


\section{Double Probe results} \label{sec:results}

The 2-point statistic of galaxy clustering can be summarized by $\{\Omega_mh^2$, $H(z)$, $D_A(z)$, $f(z)\sigma_8(z)\}$ (e.g. \cite{Chuang:2012qt}). In some studies, $\Omega_mh^2$ was not included since a strong prior had been applied. Instead of using $H(z)$ and $D_A(z)$, one uses the derived parameters $H(z)r_s/r_{s,fid}$ and $D_A(z)r_{s,fid}/r_s$ to summarize the cosmological information since these two quantities are basically uncorrelated to $\Omega_mh^2$, where $r_s$ is the sound horizon at the redshift of the drag epoch and $r_{s,fid}$ is the $r_s$ of the fiducial cosmology. In this study, $\Omega_mh^2$ is well constrained by the joint data set but we still use $H(z)r_s/r_{s,fid}$ and $D_A(z)r_{s,fid}/r_s$ because they have tighter constraints.

\cite{Wang:2007mza} showed that CMB shift parameters $(l_a, R)$, together with
$\Omega_b h^2$, provide an efficient and intuitive summary of CMB data as far as 
dark energy constraints are concerned. It is equivalent to replace
$\Omega_b h^2$ with $z_*$, the redshift to the photon-decoupling surface
\citep{Wang:2009sn}.
The CMB shift parameters are defined as \citep{Wang:2007mza}:
\ba
R \equiv \sqrt{\Omega_m H_0^2} \,r(z_*),\\
l_a \equiv \pi r(z_*)/r_s(z_*),
\ea 
and $z_{\star}$ is the redshift to the photon-decoupling surface given by CAMB \citep{Lewis:1999bs}

The angular comoving distance to an object at redshift $z$ is given by:
\be
\label{eq:r(z)}
r(z)=cH_0^{-1}\, |\Omega_k|^{-1/2} {\rm sinn}[|\Omega_k|^{1/2}\, \Gamma(z)],
\ee
which has simple relation with the angular diameter distance $D_A(z)=r(z)/(1+z)$.

In additional to the shift parameters, we include also the scalar index and 
amplitude of the power law primordial fluctuation $n_s$ and $A_s$ to summarize the CMB information.

From the measured parameters $\{\Omega_c h^2$, $\Omega_b h^2$,  $n_s$, $log(A_s)$, $\theta$, $\tau$, $\Omega_k$, $w$,$H(z)$, $D_A(z)$, $\beta(z)$, $b\sigma_8(z)$, $b(z)\}$, we derive the parameters $\{R$, $l_a$, $\Omega_b h^2$, $n_s$, $log(10^{10} A_s)$, $\Omega_k$, $H(z)r_s/r_{s,fid}$, $D_A(z)r_{s,fid}/r_s$, $f(z)\sigma_8(z)\}$ to summarize the joint data set of Planck and BOSS galaxy sample. Table\  \ref{table:fiducial_owCDM} and \ref{table:fiducial_owCDM_covar} show the measured values and their normalized covariance.
A normalized covariance matrix is defined by
\begin{equation}
N_{ij}=\frac{C_{ij}}{\sqrt{C_{ii}C_{jj}}},
\end{equation}
where $C_{ij}$ is the covariance matrix.

To conveniently compare with other measurements using CMASS sample within $0.43<z<0.7$ (we are using $0.43<z<0.75$), we extrapolated our measurements at $z=0.57$: $H(0.57)r_s/r_{s,fid}= 96.7 \pm 3.1\Hunit$ and $D_A(0.57)r_{s,fid}/r_s=1405\pm25$Mpc (see Table 9 of \citealt{Acacia}).

\begin{table}
 \begin{center}
  \begin{tabular}{c c}
     \hline 
$	f\sigma_8(0.59)	$&$	0.510	\pm	0.047	$\\
$	H(0.59)r_s/r_{s,fid}	$&$	97.9	\pm	3.1	$\\
$	D_A(0.59)r_{s,fid}/r_s	$&$	1422	\pm	25	$\\
$	f\sigma_8(0.32)	$&$	0.431	\pm	0.063	$\\
$	H(0.32)r_s/r_{s,fid}	$&$	79.1	\pm	3.3	$\\
$	D_A(0.32)r_{s,fid}/r_s	$&$	956	\pm	27	$\\
$	R	$&$	1.7430	\pm	0.0080	$\\
$	l_a	$&$	301.70	\pm	0.15	$\\
$	\Omega_bh^2	$&$	0.02233	\pm	0.00025	$\\
$	n_s	$&$	0.9690	\pm	0.0066	$\\
$	{\rm{ln}}(10^{10}A_s)	$&$	3.040	\pm	0.036	$\\
$	\Omega_k	$&$	-0.003	\pm	0.006	$\\
     \hline
 \end{tabular}
 \end{center}
\caption{Fiducial result of the double-probe approach.
The units of $H(z)$ and $D_A(z)$ are $\Hunit$ and Mpc.
}
\label{table:fiducial_owCDM}
  \end{table}

\begin{table*}\scriptsize
 \begin{center}
  \begin{tabular}{c c c c c c c c c c c c c}
     \hline \\
		&	R	&	la	&	$\Omega_bh^2$	&	$n_s$	&$	{\rm{ln}}(10^{10}A_s)	$&$	f\sigma_8(0.59)	$&$	\frac{H(0.59)}{r_{s,fid}/r_s}	$&$	\frac{D_A(0.59)}{r_s/r_{s,fid}}	$&$	f\sigma_8(0.32)	$&$\frac{H(0.32)}{r_{s,fid}/r_s}	$&$	\frac{D_A(0.32)}{r_s/r_{s,fid}}	$&$	\Omega_k		$\\	\hline
	R	&	1.0000	&	0.6534	&	-0.7271	&	-0.8787	&	-0.0352	&	-0.0620	&	-0.1675	&	-0.0059	&	-0.0237	&	-0.0271	&	0.0027	&	0.6349	\\	
	la	&	0.6534	&	1.0000	&	-0.5212	&	-0.5770	&	-0.0651	&	-0.1067	&	-0.1957	&	0.0017	&	0.0073	&	0.0174	&	-0.0211	&	0.4329	\\	
	$\Omega_bh^2$	&	-0.7271	&	-0.5212	&	1.0000	&	0.6633	&	0.1175	&	0.0525	&	0.0822	&	0.0333	&	0.1373	&	0.0566	&	0.0321	&	-0.4070	\\	
	$n_s$	&	-0.8787	&	-0.5770	&	0.6633	&	1.0000	&	0.0808	&	0.0381	&	0.1648	&	-0.0003	&	0.0285	&	0.0510	&	0.0303	&	-0.5547	\\	
$	{\rm{ln}}(10^{10}A_s)	$&	-0.0352	&	-0.0651	&	0.1175	&	0.0808	&	1.0000	&	0.0034	&	0.0391	&	0.0175	&	-0.0066	&	0.0020	&	0.0516	&	0.5915	\\	
$	f\sigma_8(0.59)	$&	-0.0620	&	-0.1067	&	0.0525	&	0.0381	&	0.0034	&	1.0000	&	0.7153	&	0.6172	&	0.1531	&	0.1535	&	-0.0333	&	0.0252	\\	
$	H(0.59)r_s/r_{s,fid}	$&	-0.1675	&	-0.1957	&	0.0822	&	0.1648	&	0.0391	&	0.7153	&	1.0000	&	0.4168	&	0.0447	&	0.0968	&	-0.0388	&	-0.0959	\\	
$	D_A(0.59)r_{s,fid}/r_s	$&	-0.0059	&	0.0017	&	0.0333	&	-0.0003	&	0.0175	&	0.6172	&	0.4168	&	1.0000	&	0.0209	&	-0.0319	&	-0.0839	&	0.0038	\\	
$	f\sigma_8(0.32)	$&	-0.0237	&	0.0073	&	0.1373	&	0.0285	&	-0.0066	&	0.1531	&	0.0447	&	0.0209	&	1.0000	&	0.6581	&	0.5250	&	0.1142	\\	
$	H(0.32)r_s/r_{s,fid}	$&	-0.0271	&	0.0174	&	0.0566	&	0.0510	&	0.0020	&	0.1535	&	0.0968	&	-0.0319	&	0.6581	&	1.0000	&	0.3168	&	0.1165	\\	
$	D_A(0.32)r_{s,fid}/r_s	$&	0.0027	&	-0.0211	&	0.0321	&	0.0303	&	0.0516	&	-0.0333	&	-0.0388	&	-0.0839	&	0.5250	&	0.3168	&	1.0000	&	0.0835	\\	
$	\Omega_k	$&	0.6349	&	0.4329	&	-0.4070	&	-0.5547	&	0.5915	&	0.0252	&	-0.0959	&	0.0038	&	0.1142	&	0.1165	&	0.0835	&	1.0000	\\		
     \hline
 \end{tabular}
 \end{center}
\caption{Normalized covariance matrix of the fiducial result from the double-probe approach.}
\label{table:fiducial_owCDM_covar}
  \end{table*}

\section{constrain parameters of given dark energy models with double-probe results} 
\label{sec:use}
In this section, we describe the steps to combine our results with other data sets assuming some dark energy models. 
For a given model and cosmological parameters, one can compute 
$\{R$, $l_a$, $\Omega_b h^2$, $n_s$, $log(10^10 A_s)$, $\Omega_k$, $H(z)r_s/r_{s,fid}$, $D_A(z)r_{s,fid}/r_s$, $f(z)\sigma_8(z)\}$.
one can take the covariance matrices, $M_{ij,\textrm{CMB+galaxy}}$, of these 12 parameters (galaxy sample are divided in two redshift bins). Then, $\chi^2_{\textrm{CMB+galaxy}}$ can be computed by
\begin{equation}
 \chi^2_{\textrm{CMB+galaxy}}=\Delta_{\textrm{CMB+galaxy}}M_{ij,\textrm{CMB+galaxy}}^{-1}\Delta_{\textrm{CMB+galaxy}},
\end{equation}
where 
\begingroup
\everymath{\scriptstyle}
\small
\begin{equation}
 \Delta_{\textrm{CMB+galaxy}}=
\left(\begin{array}{c}
f\sigma_8(0.59)	- 0.510\\
H(0.59)r_s/r_{s,fid}-97.9\\
D_A(0.59)r_{s,fid}/r_s-	1422\\
f\sigma_8(0.32)-	0.431\\
H(0.32)r_s/r_{s,fid}-	79.1\\
D_A(0.32)r_{s,fid}/r_s-	956\\
R	-	1.7430\\
l_a	-301.70\\
\Omega_bh^2-0.02233\\
n_s	-0.9690\\
{\rm{ln}}(10^{10}A_s)-	3.040\\
\Omega_k-0.003\\
\end{array}\right),
\end{equation}
\endgroup
where the angular diameter distance $D_A(z)$ is given by: 
\be
\label{eq:da}
 D_A(z)=(1+z)cH_0^{-1}\, |\Omega_k|^{-1/2} {\rm sinn}[|\Omega_k|^{1/2}\, \Gamma(z)],
\ee
\begin{equation*}
\mbox{where }\Gamma(z)=\int_0^z\frac{dz'}{E(z')}, \mbox{ and } E(z)=H(z)/H_0,
\end{equation*}
and ${\rm sinn}(x)=\sin(x)$, $x$, $\sinh(x)$ for 
$\Omega_k<0$, $\Omega_k=0$, and $\Omega_k>0$ respectively;
and the expansion rate the universe $H(z)$ is given by
\ba
H(z) = \nonumber
\ea
\ba
H_0 \sqrt{\Omega_m (1+z)^3 +\Omega_r (1+z)^4 +\Omega_k (1+z)^2 
+ \Omega_X X(z)},\label{eq:H}
\ea
where $\Omega_m+\Omega_r+\Omega_k+\Omega_X=1$, and
the dark energy density function $X(z)$ is defined as
\be
X(z) \equiv \frac{\rho_X(z)}{\rho_X(0)}.
\ee
$f$ is defined in relation to the linear growth factor $D(\tau)$ in the usual way as
\begin{equation}\label{eq:f}
f = \frac{d\ln D(\tau)}{d\ln a} = \frac{1}{\mathcal{H}}\frac{d\ln D(\tau)}{d\tau},
\end{equation}
where $D$ is the growing solution to the second order differential equation writen in comoving coordinates
\begin{equation}
\frac{d^2D(\tau)}{d\tau^2} + \mathcal{H}\frac{dD(\tau)}{d\tau} = \frac{3}{2}\Omega_m(\tau)\mathcal{H}^2(\tau)D(\tau).
\end{equation}
We will be writing $\sigma(z,R)$ as:
\begin{equation}
\sigma^2(z,R) = \frac{1}{(2\pi)^3}\int d^3k W^2(kR) P(k,z)
\end{equation}
with 
\begin{equation}
W(kR) = \frac{3}{(kR)^3} [\sin(kR) - kR\cos(kR)]
\end{equation}
being the top-hat window function. Thus
\begin{equation}\label{eq:s8}
 \sigma_8(z)=\sigma(z,R=8Mpc/h).
\end{equation}
In this way, one just need to compute linear theory to get
$\chi^2_{\textrm{CMB+galaxy}}$ to reproduce and combine CMB plus galaxy information. These equations assume no impact from massive 
neutrinos, mainly working for the cases of massless or approximately massless neutrinos. When including neutrino species with a given mass
one needs to solve the full Boltzmann hierarchy as shown in \cite{Ma:1995ey,Lewis:2002nc}.

Table \ref{table:assume_DEmodel} lists the constraints on the parameters of different dark energy models obtained using our double-probe measurements.
The results show no tension with the flat $\Lambda$CDM cosmological paradigm.

\begin{table*}
 \begin{center}
  \begin{tabular}{c c c c c c c}
     \hline 
	&	$\Omega_m$	&			$H_0$	&			$\sigma_8$	&			$\Omega_k$	&			$w$ or $w_0$	&			$w_a$			\\	\hline
$\Lambda$CDM	& $	0.304	\pm	0.009	$&$	68.2	\pm	0.7	$&$	0.806	\pm	0.014	$&$	0			$&$	-1			$&$	0			$\\	
o$\Lambda$CDM	& $	0.303	\pm	0.010	$&$	68.6	\pm	0.9	$&$	0.810	\pm	0.015	$&$	0.002	\pm	0.003	$&$	-1			$&$	0			$\\	
$w$CDM	& $	0.299	\pm	0.013	$&$	69.0	\pm	1.5	$&$	0.815	\pm	0.020	$&$	0			$&$	-1.04	\pm	0.06	$&$	0			$\\	
o$w$CDM	& $	0.302	\pm	0.014	$&$	68.7	\pm	1.5	$&$	0.811	\pm	0.021	$&$	0.002	\pm	0.003	$&$	-1.00	\pm	0.07	$&$	0			$\\	
$w_0w_a$CDM	& $	0.313	\pm	0.020	$&$	67.6	\pm	2.0	$&$	0.817	\pm	0.016	$&$	0			$&$	-0.84	\pm	0.22	$&$	-0.66	\pm	0.68	$\\	
o$w_0w_a$CDM	& $	0.313	\pm	0.020	$&$	67.6	\pm	2.2	$&$	0.815	\pm	0.016	$&$	0.000	\pm	0.004	$&$	-0.85	\pm	0.24	$&$	-0.61	\pm	0.80	$\\	
\hline
\end{tabular}
 \end{center}
\caption{Constraints on cosmological parameters obtained by using our results assuming dark energy models (see Sec. \ref{sec:use}).}
\label{table:assume_DEmodel}
  \end{table*}

\section{Full-likelihood analysis Fixing Dark Energy Models}
\label{sec:full_run}
\begin{figure*}
\centering
\subfigure{\includegraphics[width=1 \columnwidth,clip,angle=0]{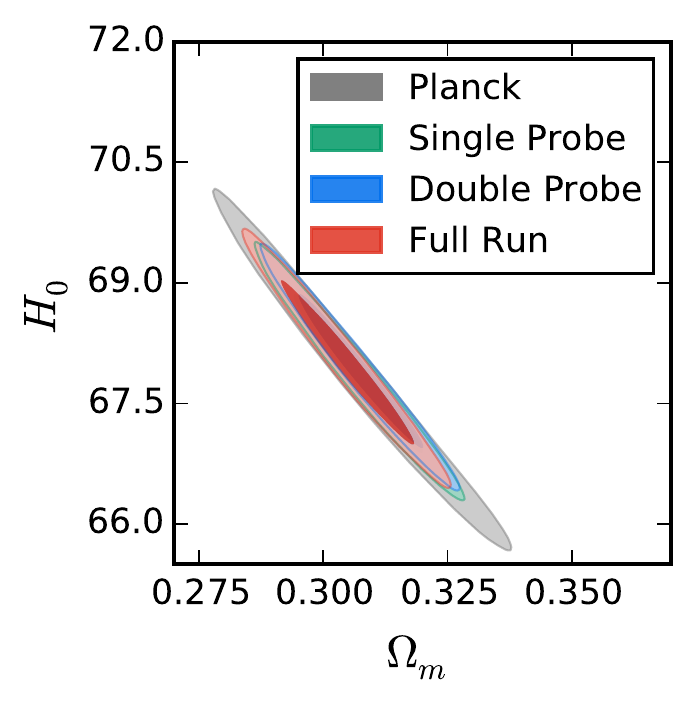}}
\subfigure{\includegraphics[width=1 \columnwidth,clip,angle=0]{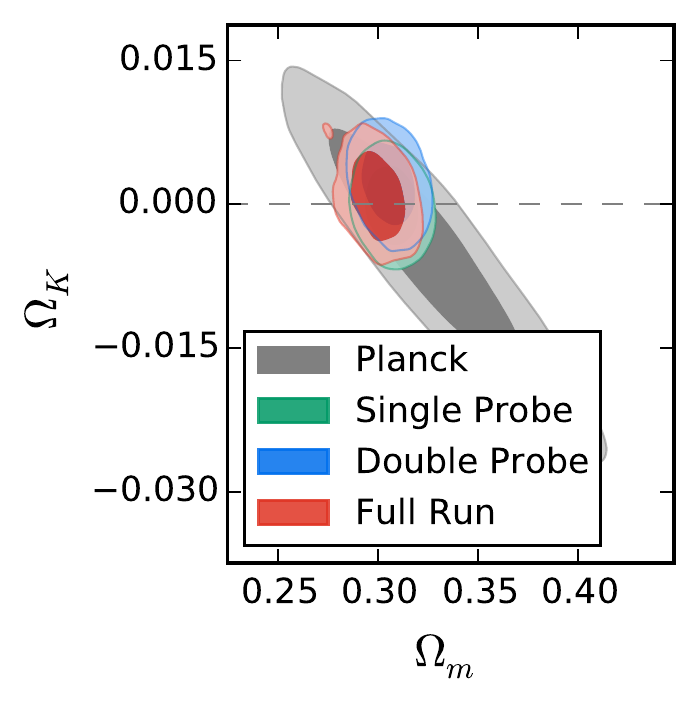}}
\caption{
Left panel: 2D marginalized
  contours for $68\%$ and $95\%$ confidence levels for $\Omega_m$ and $H_0$ ($\Lambda$CDM model assumed)
from Planck-only (gray), derived using double probe measurements (blue), full -likelihood analysis with joint data (red; labeled as "Full Run"), and Planck+single probe measurements (green).
Right panel: 2D marginalized
  contours for $68\%$ and $95\%$ confidence level for $\Omega_m$ and $\Omega_k$ (o$\Lambda$CDM model assumed).
One can see that the latter three measurements are consistent with each other.
}
\label{fig:lcdm}
\end{figure*}

\begin{figure*}
\centering
\subfigure{\includegraphics[width=1 \columnwidth,clip,angle=0]{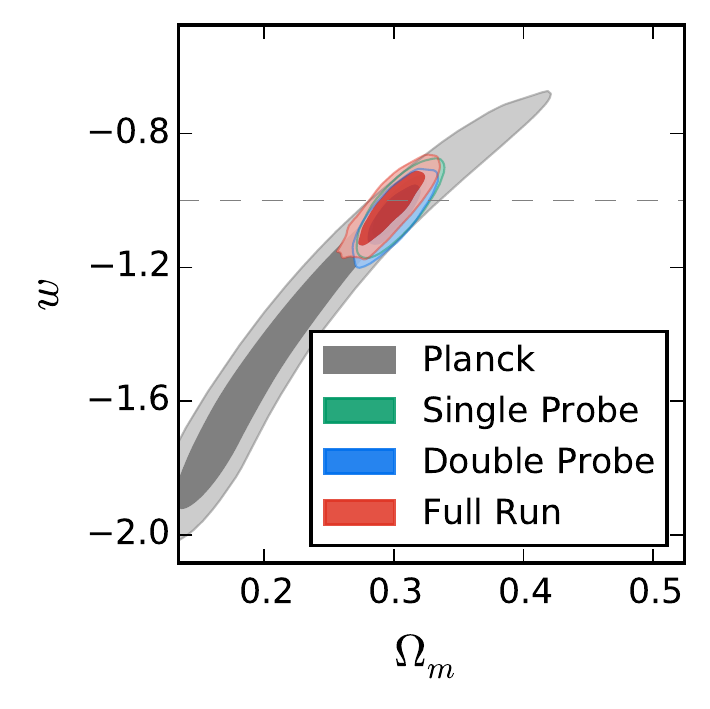}}
\subfigure{\includegraphics[width=1 \columnwidth,clip,angle=0]{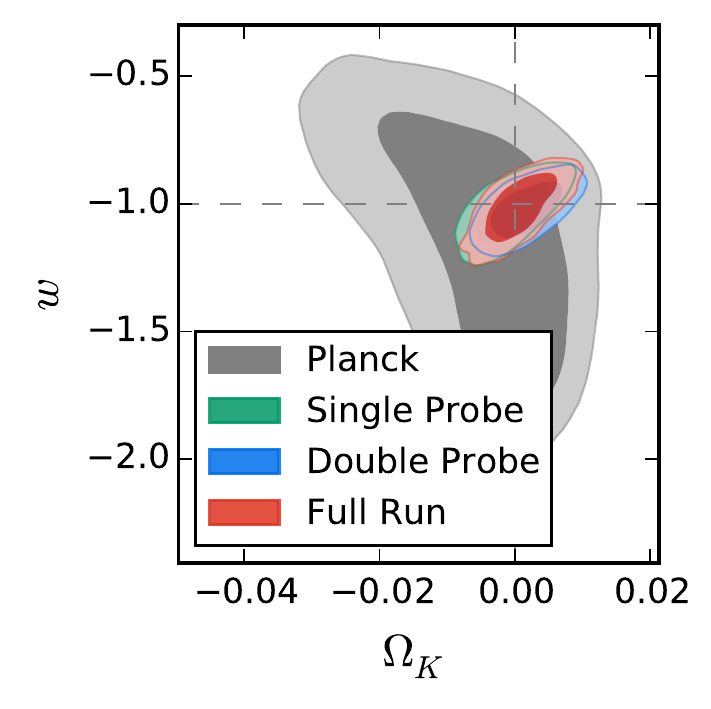}}
\caption{
Left panel: 2D marginalized
  contours for $68\%$ and $95\%$ confidence level for $\Omega_m$ and $w$ ($w$CDM model assumed)
from Planck-only (gray), derived using double probe measurements (blue),  full -likelihood analysis with joint data (red; labeled as "Full Run"), and Planck+single probe measurements (green).
Right panel: 2D marginalized
  contours for $68\%$ and $95\%$ confidence level for $\Omega_k$ and $w$ (o$w$CDM model assumed).
One can see that the latter three measurements are consistent with each other.
}
\label{fig:wcdm}
\end{figure*}

\begin{figure*}
\centering
\subfigure{\includegraphics[width=1 \columnwidth,clip,angle=0]{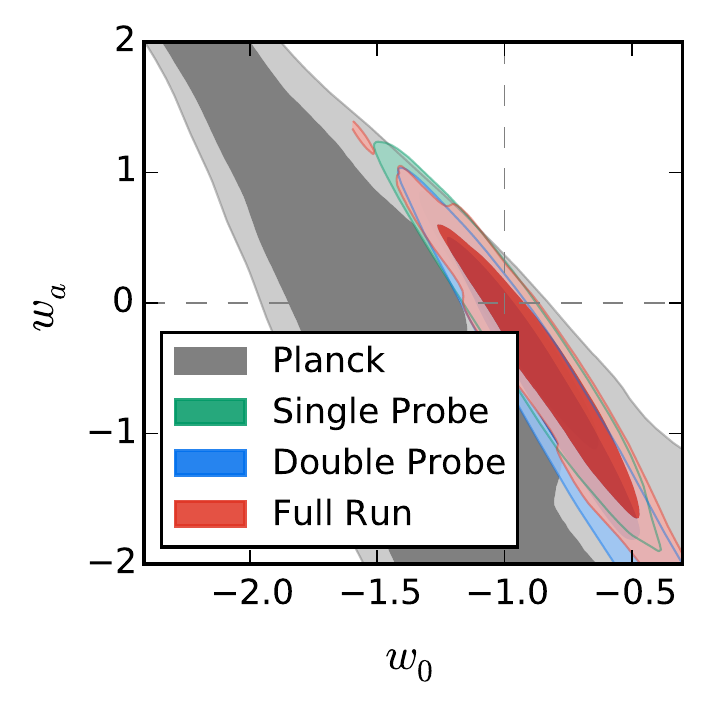}}
\subfigure{\includegraphics[width=1 \columnwidth,clip,angle=0]{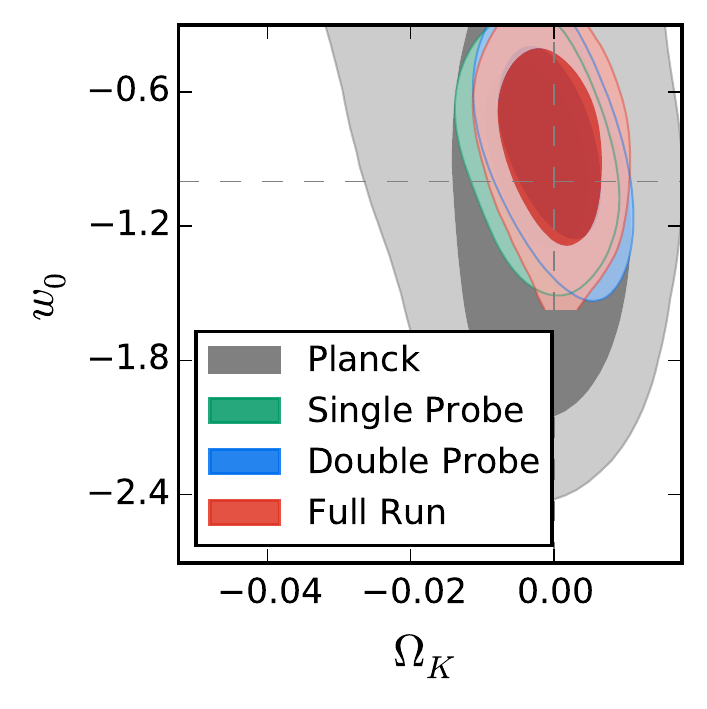}}
\caption{
Left panel: 2D marginalized
  contours for $68\%$ and $95\%$ confidence level for $w_0$ and $w_a$ ($w_0w_a$CDM model assumed)
from Planck-only (gray), derived using double probe measurements (blue), full -likelihood analysis with joint data (red; labeled as "Full Run"), and Planck+single probe measurements (green).
Right panel: 2D marginalized
  contours for $68\%$ and $95\%$ confidence level for $\Omega_k$ and $w_0$ (o$w_0w_a$CDM model assumed).
One can see that the latter three measurements are consistent with each other.
}
\label{fig:w0wa}
\end{figure*}

To validate our double-probe methodology, we perform the full-likelihood MCMC analyses with fixing dark energy models. The main difference of this approach comparing our double-probe analysis is that it has been given a dark energy model at first place. Opposite to the double probe approach, one cannot use the results from the full-likelihood analysis to derive the constraints for the parameters of other dark energy models.
Since the dark energy model is fixed, the quantities, $\{H(z)$, $D_A(z)$, $\beta(z)$, $b\sigma_8(z)\}$, would be determined by the input parameters, $\{\Omega_c h^2$, $\Omega_b h^2$,  $n_s$, $log(A_s)$, $\theta$, $\tau$, $\Omega_k$, $w\}$, as shown in Eq. \ref{eq:da}, \ref{eq:H}, \ref{eq:f}, and \ref{eq:s8}.
We show the results in Table \ref{table:fix_DEmodel}. In Fig. \ref{fig:lcdm}, \ref{fig:wcdm} and \ref{fig:w0wa}, we compare these results with our double-probe approach and the single-probe approach (\cite{Chuang16}; companion paper).  We find very good agreement among these three approaches.
Note that deriving the dark energy model constraints from our double-probe measurements is much faster than the full run. For example, using the same machine, it takes $\sim2.5$ hours to obtain the constraints for $\Lambda$CDM using double-probe measurements, but takes 6 days to reach similar convergence for the full likelihood MCMC analysis (slower with a factor of 60).

Up to this point we have introduced two methodologies for extracting cosmological information, the double-probe method and
a full likelihood analysis. Moreover, we are comparing these results with a third methodology already introduced in Chuang et al. 2016
also called single-probe analysis combined with CMB. We show here motivations for the use of each of them:

\begin{itemize}
 \item Double-probe: Joint fit to LSS data and CMB constraining the full set of cosmological parameters without the need of extra knowledge
 on the priors. This methodology allow us to test on the prior information content assumed by other probes and give us the tool to have 
 a dark energy independent measurements from LSS and CMB combined.
 \item Full fit: Fit of cosmological parameter set to LSS and CMB data,
requiring an assumption of a dark energy model (i.e. not going through $D_A$, $H$ and
$f\sigma_8$ as intermediate parameters) from the beginning. This methodology provides a tool to check the information content of the data
and we take it to be the answer to recover from other methodologies as it does not have extra assumptions appart from the dark energy
model.
 \item Single-probe+CMB: Likelihoods are determined from the BOSS measurements of
$\{D_Ar_s^{fid}/r_s$, $Hr_s/r_s^{fid}$, $f\sigma_8$, $\Omega_mh^2\}$ together with Planck data. This methodology provides, in its first 
step, measurements 
of large scale structure independent of CMB data, thus showing as a good tool to test possible tensions between data sets.
\end{itemize}

\begin{table*}
 \begin{center}
  \begin{tabular}{c c c c c c c}
     \hline 
	&	$\Omega_m$	&			$H_0$	&			$\sigma_8$	&			$\Omega_k$	&			$w$ or $w_0$	&			$w_a$			\\	\hline
$\Lambda$CDM	& $	0.305	\pm	0.008	$&$	68.0	\pm	0.6	$&$	0.812	\pm	0.009	$&$	0			$&$	-1			$&$	0			$\\	
o$\Lambda$CDM	& $	0.300	\pm	0.009	$&$	68.6	\pm	1.0	$&$	0.816	\pm	0.010	$&$	0.001	\pm	0.003	$&$	-1			$&$	0			$\\	
$w$CDM	& $	0.298	\pm	0.015	$&$	68.8	\pm	1.8	$&$	0.818	\pm	0.017	$&$	0			$&$	-1.02	\pm	0.07	$&$	0			$\\	
o$w$CDM	& $	0.298	\pm	0.017	$&$	68.8	\pm	1.8	$&$	0.818	\pm	0.018	$&$	0.001	\pm	0.003	$&$	-1.01	\pm	0.08	$&$	0			$\\	
$w_0w_a$CDM	& $	0.311	\pm	0.022	$&$	67.4	\pm	2.3	$&$	0.808	\pm	0.020	$&$	0			$&$	-0.85	\pm	0.23	$&$	-0.51	\pm	0.67	$\\	
o$w_0w_a$CDM	& $	0.309	\pm	0.025	$&$	67.8	\pm	3.0	$&$	0.810	\pm	0.024	$&$	0.000	\pm	0.004	$&$	-0.86	\pm	0.26	$&$	-0.50	\pm	0.73	$\\			
\hline
\end{tabular}
 \end{center}
\caption{Constraints on cosmological parameters from full-likelihood MCMC analysis of the joint data set (see Sec. \ref{sec:full_run}).}
\label{table:fix_DEmodel}
  \end{table*}

\input{mnu}

\section{Summary} 
\label{sec:conclusion}
 In this work we have studied and compared three different ways of extracting cosmological information from the combined data sets 
 of Planck2015 and BOSS final data release (DR12) having great care in avoiding imposing priors on cosmological parameters when combining these data. 
 
First, we have extracted the dark-energy-model-independent cosmological constraints from the joint data sets of Baryon Oscillation Spectroscopic Survey (BOSS) galaxy sample and Planck cosmic microwave background (CMB) measurement.
We measure the mean values and covariance matrix of $\{R$, $l_a$, $\Omega_b h^2$, $n_s$, $log(A_s)$, $\Omega_k$, $H(z)$, $D_A(z)$, $f(z)\sigma_8(z)\}$, which give an efficient summary of Planck data and 2-point statistics from BOSS galaxy sample (see Table \ref{table:fiducial_owCDM}). We called this methodology as "double-probe" approach since it combines two data sets to minimize the priors needed for the cosmological parameters.
We found that double probe measurements are insensitive to the assumption of neutrino mass (fixed or not). But, the parameter $R$ should be replaced by $\Omega_{bc}h^2$ while having $\Sigma m_\nu$ to be free.

Second, we performed the full-likelihood-analysis from the joint data set of Planck and BOSS assuming some simple dark energy models. By comparing these results with the ones from double-probe approach, we have demonstrated that the double-probe approach provides robust cosmological parameter constraints which can be conveniently used to study dark energy models.
Using our results, we obtain 
$\Omega_m=0.304\pm0.009$, $H_0=68.2\pm0.7$, and $\sigma_8=0.806\pm0.014$ assuming $\Lambda$CDM; 
$\Omega_k=0.002\pm0.003$ assuming oCDM; $w=-1.04\pm0.06$ assuming $w$CDM; 
$\Omega_k=0.002\pm0.003$ and $w=-1.00\pm0.07$ assuming o$w$CDM;
and $w_0=-0.84\pm0.22$ and $w_a=-0.66\pm0.68$ assuming $w_0w_a$CDM.
The results show no tension with the flat $\Lambda$CDM cosmological paradigm.
Note that deriving the dark energy model constraints from our double-probe measurements is much faster than the full run. For example, it takes $\sim2.5$ hours to obtain the constraints for $\Lambda$CDM using double-probe measurements, but takes 6 days to reach similar convergence for the full MCMC run (slower with a factor of 60).

We have extended our study to measure the sum of neutrino mass using different methodologies including double probe analysis (introduced in this study), full-likelihood analysis, and single probe analysis. 
We found that double probe has weaker constraint on the neutrino mass since it does not include the constraining power on the neutrino mass from Planck data.
While including lensing information, we have performed the analyses with varying $A_{\rm L}$ or fixing $A_{\rm L}=1$. We found that varying $A_{\rm L}$ would shift the $\Sigma m_\nu$ to a larger value. 
From the full-likelihood analysis with varying $A_{\rm L}$, we obtained
$\Sigma m_\nu=0.17^{+0.08}_{-0.13}$ assuming $\Lambda$CDM;
$\Sigma m_\nu=0.34^{+0.17}_{-0.22}$ assuming o$\Lambda$CDM; 
$\Sigma m_\nu=0.33^{+0.16}_{-0.18}$ assuming $w$CDM;
$\Sigma m_\nu=0.44^{+0.23}_{-0.22}$ assuming o$w$CDM.
We found $\sim2\sigma$ detection of $\Sigma m_\nu$ when allowing $w$ and $\Omega_k$ to be free. 

In addition, when performing the full-likelihood analysis, we found that the overall shape of correlation function contributed to the detection of neutrino mass significantly. However, since we do not have high confidence on the overall shape because of the potential observational systematics, we removed the overall shape information to be conservative. The numbers provided above have been obtained without the overall shape information. Our study have shown that one should be cautious to the impact of observational systematics when constraining neutrino mass using the large scale structure measurements.

\section*{Acknowledgement}
M.P.I. would like to thank Denis Tramonte and Rafael Rebolo for useful discussions.
M.P.I. and C.C. thanks David Hogg, Savvas Nesseris, and Yun Wang for
useful discussions.
C.C. and F.P. acknowledge support from the Spanish MICINN’s Consolider-Ingenio 2010 Programme under grant MultiDark CSD2009-00064 and AYA2010-21231-C02-01 grant.
C.C. was also supported by the Comunidad de Madrid under grant HEPHACOS S2009/ESP-1473. C.C. was supported as a MultiDark fellow.
M.P.I. acknowledges support from MINECO under the grant AYA2012-39702-C02-01.
G.R. is supported by the National Research Foundation of Korea (NRF) through NRF-SGER 2014055950 funded by the Korean Ministry of Education, Science and Technology (MoEST), and by the faculty research fund of Sejong University in 2016.

We acknowledge the use of 
the CURIE supercomputer at Tr\`es Grand Centre de calcul du CEA in France  through the French participation into the PRACE research infrastructure,
the SuperMUC supercomputer at Leibniz Supercomputing Centre of the Bavarian Academy of Science in Germany,
the TEIDE-HPC (High Performance Computing) supercomputer in Spain,
and the Hydra cluster at Instituto de F\'{\i}sica Te\'orica, (UAM/CSIC) in Spain.

Funding for SDSS-III has been provided by the Alfred P. Sloan Foundation, the Participating Institutions, the National Science Foundation, 
and the U.S. Department of Energy Office of Science. The SDSS-III web site is http://www.sdss3.org/.

SDSS-III is managed by the Astrophysical Research Consortium for the Participating Institutions of the SDSS-III Collaboration including 
the University of Arizona, the Brazilian Participation Group, Brookhaven National Laboratory, Carnegie Mellon University, University of Florida, 
the French Participation Group, the German Participation Group, Harvard University, the Instituto de Astrofisica de Canarias, 
the Michigan State/Notre Dame/JINA Participation Group, Johns Hopkins University, Lawrence Berkeley National Laboratory, Max Planck Institute for Astrophysics, 
Max Planck Institute for Extraterrestrial Physics, New Mexico State University, New York University, Ohio State University, 
Pennsylvania State University, University of Portsmouth, Princeton University, the Spanish Participation Group, University of Tokyo, University of Utah, 
Vanderbilt University, University of Virginia, University of Washington, and Yale University. 

\bibliography{doubleprobe} 


\label{lastpage}

\end{document}

%% file: mnu.tex
\section{measurements of neutrino mass}
\label{sec:mnu}
In this section, we will focus on measuring the sum of the neutrino mass $\Sigma m_\nu$ using different methodologies described in previous sections.
First, we repeat the double-probe analysis described in Sec. \ref{sec:mcmc} with an additional free parameter, $\Sigma m_\nu$, and present the constraints on cosmological parameters.
Second, we repeat the MCMC analysis with the full likelihood of joint data set described in Sec. \ref{sec:full_run} and find that the full shape measurement of the monopole of the galaxy 2-point correlation function introduces some detection of neutrino mass. However, since the monopole measurement is sensitive to the observational systematics, we provide another set of cosmological constraints by removing the full shape information.
Third, we also obtain the constraint of $\Sigma m_\nu$ using the single probe measurement provided by Chuang et al. (companion paper).

\subsection{measuring neutrino mass using double probe}
\label{sec:mnu_double}
Note first that for the study of $m_{\nu}$, we replace $R=\sqrt{\Omega_m H_0^2}\,r(z_*)$ with $\Omega_{bc}h^2=\Omega_bh^2+\Omega_ch^2$ (e.g. see \cite{Aubourg:2014yra}), since $R$ depends directly on 
$\Omega_{\nu}$. 
Thus, we use the
following set of parameters from the double probe analysis while measuring neutrino mass, 
$\{\Omega_{bc}h^2$, $l_a$, $\Omega_b h^2$, $n_s$, $log(A_s)$, $\Omega_k$, $H(z)$, $D_A(z)$, $f(z)\sigma_8(z)\}$.

We repeat the analysis described in Sec. \ref{sec:mcmc}, but here we set $\Sigma m_\nu$, to be free instead of setting it to $0.06$ eV. The results are shown in Table \ref{table:result_owCDM_mnu} and \ref{table:result_owCDM_mnu_covar}.

\begin{figure}
\centering
\includegraphics[width=1 \columnwidth,clip,angle=0]{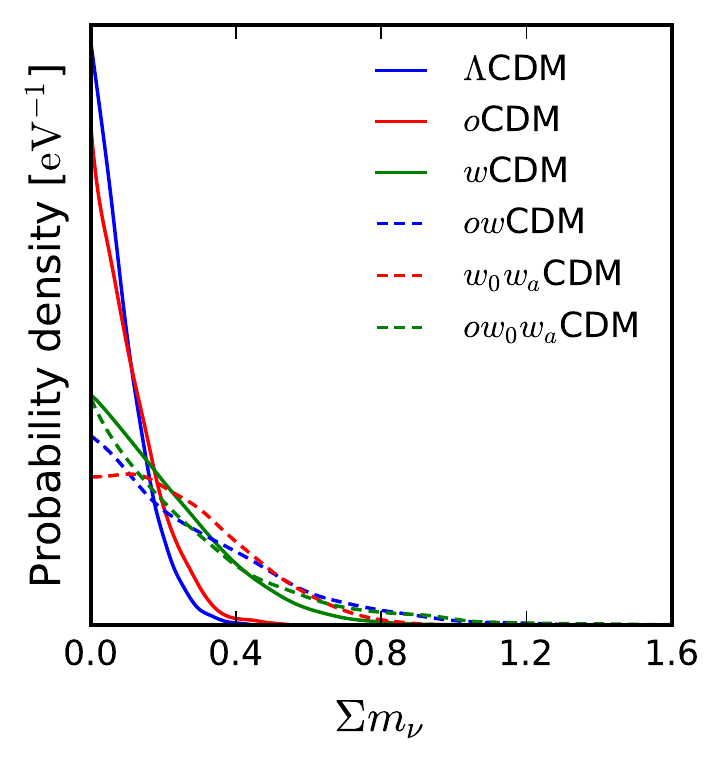}
\caption{
Probability density for $\Sigma m_\nu$
from double-probe measurements using the covariance matrix with free parameter $\Sigma m_\nu$ (see Sec. \ref{sec:mnu_double} and Table \ref{table:assume_DEmodel_mnu_vary}). 
}
\label{fig:mnu_doubleprobe}
\end{figure}

As described in Sec. \ref{sec:use}, one can constrain the parameters of given dark energy models using Table 
\ref{table:result_owCDM_mnu} and \ref{table:result_owCDM_mnu_covar}. 
Table \ref{table:assume_DEmodel_mnu_vary}  presents the cosmological parameter constraints 
assuming some simple dark energy models.
Figure \ref{fig:mnu_doubleprobe} shows the probability density for $\Sigma m_\nu$ for different dark energy models. Our measurements of $\Sigma m_\nu$ using double probe approach are consistent with zero. The upper limit (68\% confidence level) varys from 0.1 to 0.35 eV depending on dark energy model.

\begin{figure}
\centering
\includegraphics[width=1 \columnwidth,clip,angle=0]{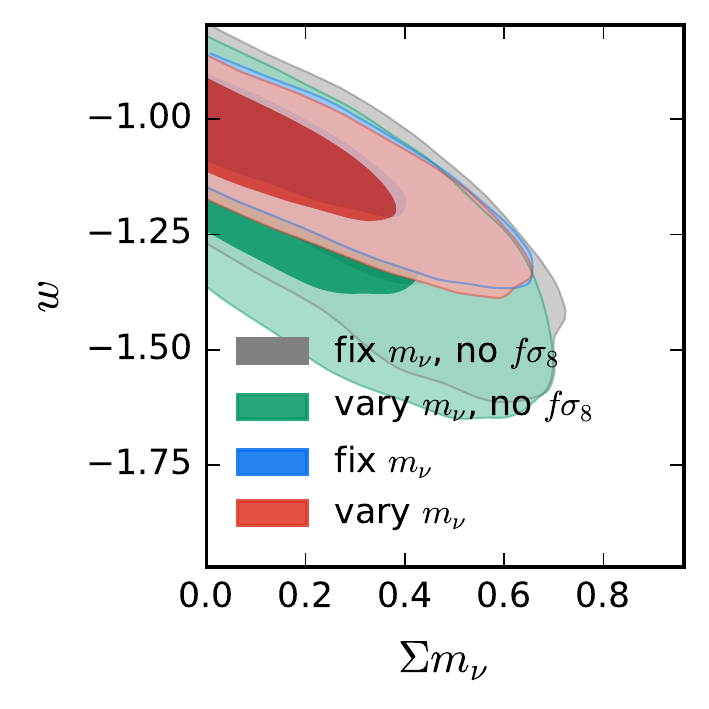}
\caption{
Comparison of 2D contours for $68\%$ and $95\%$ confidence level on $\Sigma m_\nu$ and $w$
from the double probe methodology using covariance matrix from first step varying and fixing neutrinos.
One can see that the constraints are insensitive to the assumption of $\Sigma m_\nu$. We also show the results from double probe measurement excluding $f(z)\sigma_8(z)$. One can see that $f(z)\sigma_8(z)$ improve the constraint on $w$ but not $\Sigma m_\nu$.
}
\label{fig:mnu_double_probe_mnucovmat_nofs8}
\end{figure}

In addition, we also derive the cosmological constraints by using the results with fixed $\Sigma m_\nu$, i.e. Table 
\ref{table:fiducial_owCDM} and \ref{table:fiducial_owCDM_covar} with $R$ replaced by $\Omega_{bc}h^2$. Different from Table \ref{table:assume_DEmodel} (see Sec. \ref{sec:use}), 
we include $\Sigma m_\nu$ as one of the parameters to be constrained. The results are shown in Table \ref{table:assume_DEmodel_mnu_fixed}.
We find that the results are very similar to Table \ref{table:assume_DEmodel_mnu_vary}, which showing our double probe measurements are insensitive to the $\Sigma m_\nu$ assumption.
Fig. \ref{fig:mnu_double_probe_mnucovmat_nofs8} shows this point in a clear way by comparing the 2D contours when including a 
covariance matrix varying 
$\Sigma m_\nu$ (using Table \ref{table:result_owCDM_mnu} and \ref{table:result_owCDM_mnu_covar}) or fixing $\Sigma m_\nu$ (using Table \ref{table:fiducial_owCDM} and \ref{table:fiducial_owCDM_covar}). We see that they lie on top of each other.
Moreover, Fig. \ref{fig:mnu_double_probe_mnucovmat_nofs8} also exhibit the constraint given by $f\sigma_8$ on the $\Sigma m_\nu$ and 
$w$ parameters. We find the constraint on $w$ become tighter while that in $\Sigma m_\nu$ stays the same when including the $f\sigma_8$ constraint. 
This is a good news for future experiments as their power on the neutrino constraint would not highly rely on the growth rate measurements which are more sensitive to the observational systematics.

\begin{figure}
\centering
\includegraphics[width=1 \columnwidth,clip,angle=0]
{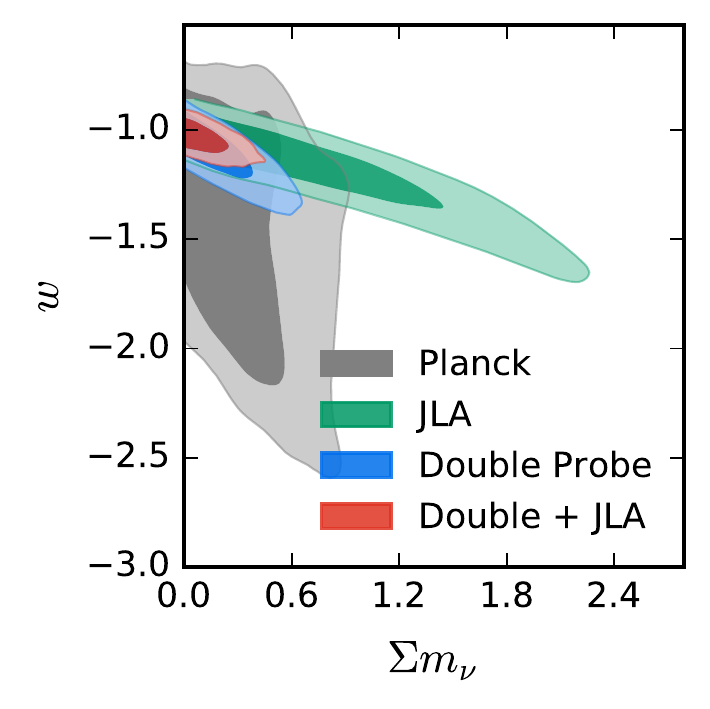}
\caption{
2D marginalized
  contours for $68\%$ and $95\%$ confidence level for $w$ and $\Sigma m_\nu$ ($w$CDM model assumed)
from Planck-only (gray), double-probe (blue), JLA (green), and double probe + JLA (red).
}
\label{fig:w_mnu_SN}
\end{figure}

Furthermore, we have also checked the impact of adding supernovae Ia (SNIa) data, dubbed Joint Light-curve Analysis (JLA) \citep{Betoule:2014frx}
and find that the upper limit of $\Sigma m_\nu$ decrease because SNIa breaks the degeneracy of the constraint from Planck+BOSS 
(see Fig. \ref{fig:w_mnu_SN}). In this way, we can get tighter constraints on the upper limit by including SNIa data.

\begin{table}
 \begin{center}
  \begin{tabular}{c c}
     \hline 
$	f\sigma_8(0.59)	$&$	0.495	\pm	0.051	$\\
$	H(0.59)r_s/r_{s,fid}	$&$	97.5	\pm	3.2	$\\
$	D_A(0.59)r_{s,fid}/r_s	$&$	1419	\pm	27	$\\
$	f\sigma_8(0.32)	$&$	0.431	\pm	0.066	$\\
$	H(0.32)r_s/r_{s,fid}	$&$	78.9	\pm	3.6	$\\
$	D_A(0.32)r_{s,fid}/r_s	$&$	964	\pm	26	$\\
$	\Omega_{bc}h^2	$&$	0.1413	\pm	0.0022	$\\
$	l_a	$&$	301.75	\pm	0.14	$\\
$	\Omega_bh^2	$&$	0.02209	\pm	0.00025	$\\
$	n_s	$&$	0.9639	\pm	0.0068	$\\
$	{\rm{ln}}(10^{10}A_s)	$&$	3.062	\pm	0.040	$\\
$	\Omega_k	$&$	-0.009	\pm	0.006	$\\
     \hline
 \end{tabular}
 \end{center}
\caption{Results of double-probe analysis obtained with varying $\Sigma m_\nu$. The units of $H(z)$ and $D_A(z)$ are $\Hunit$ and Mpc (see Sec. \ref{sec:mnu_double}).}
\label{table:result_owCDM_mnu}
  \end{table}

\begin{table*}\scriptsize
 \begin{center}
  \begin{tabular}{c c c c c c c c c c c c c}
     \hline \\
		&	$\Omega_{bc}h^2$	&	$l_a$	&	$\Omega_bh^2$	&	$n_s$	&$	{\rm{ln}}(10^{10}A_s)	$&$	f\sigma_8(0.59)	$&$	\frac{H(0.59)}{r_{s,fid}/r_s}	$&$	\frac{D_A(0.59)}{r_s/r_{s,fid}}	$&$	f\sigma_8(0.32)	$&$	\frac{H(0.32)}{r_{s,fid}/r_s}	$&$	\frac{D_A(0.32)}{r_s/r_{s,fid}}	$&$	\Omega_k	$\\	\hline
$	\Omega_{bc}h^2	$&	1.0000	&	0.4607	&	-0.6377	&	-0.8376	&	0.0145	&	0.0075	&	0.0536	&	0.0672	&	-0.0870	&	0.0317	&	0.0049	&	0.3794	\\	
$	l_a	$&	0.4607	&	1.0000	&	-0.4977	&	-0.5042	&	-0.0470	&	0.0201	&	-0.0525	&	0.0043	&	-0.0216	&	0.0765	&	0.0912	&	0.2919	\\	
	$\Omega_bh^2$	&	-0.6377	&	-0.4977	&	1.0000	&	0.7188	&	-0.0241	&	-0.0016	&	-0.0625	&	-0.0879	&	0.0692	&	0.0299	&	0.0149	&	-0.2708	\\	
	$n_s$	&	-0.8376	&	-0.5042	&	0.7188	&	1.0000	&	0.0475	&	-0.0131	&	-0.0591	&	-0.0499	&	0.0717	&	0.0268	&	-0.0686	&	-0.2894	\\	
$	{\rm{ln}}(10^{10}A_s)	$&	0.0145	&	-0.0470	&	-0.0241	&	0.0475	&	1.0000	&	0.0095	&	-0.0352	&	-0.0065	&	0.0773	&	0.0225	&	0.0053	&	0.5576	\\	
$	f\sigma_8(0.59)	$&	0.0075	&	0.0201	&	-0.0016	&	-0.0131	&	0.0095	&	1.0000	&	0.6546	&	0.5223	&	0.2427	&	0.2074	&	0.0634	&	0.1538	\\	
$	H(0.59)r_s/r_{s,fid}	$&	0.0536	&	-0.0525	&	-0.0625	&	-0.0591	&	-0.0352	&	0.6546	&	1.0000	&	0.3777	&	0.0586	&	0.0615	&	0.0015	&	-0.0025	\\	
$	D_A(0.59)r_{s,fid}/r_s	$&	0.0672	&	0.0043	&	-0.0879	&	-0.0499	&	-0.0065	&	0.5223	&	0.3777	&	1.0000	&	-0.0598	&	0.0272	&	-0.0474	&	-0.0578	\\	
$	f\sigma_8(0.32)	$&	-0.0870	&	-0.0216	&	0.0692	&	0.0717	&	0.0773	&	0.2427	&	0.0586	&	-0.0598	&	1.0000	&	0.6531	&	0.4819	&	0.1487	\\	
$	H(0.32)r_s/r_{s,fid}	$&	0.0317	&	0.0765	&	0.0299	&	0.0268	&	0.0225	&	0.2074	&	0.0615	&	0.0272	&	0.6531	&	1.0000	&	0.1686	&	0.1165	\\	
$	D_A(0.32)r_{s,fid}/r_s	$&	0.0049	&	0.0912	&	0.0149	&	-0.0686	&	0.0053	&	0.0634	&	0.0015	&	-0.0474	&	0.4819	&	0.1686	&	1.0000	&	0.0049	\\	
$	\Omega_k	$&	0.3794	&	0.2919	&	-0.2708	&	-0.2894	&	0.5576	&	0.1538	&	-0.0025	&	-0.0578	&	0.1487	&	0.1165	&	0.0049	&	1.0000	\\	
     \hline
 \end{tabular}
 \end{center}
\caption{Correlation matrix of the double-probe measurements obtained with varying $\Sigma m_\nu$ (corresponding to Table \ref{table:result_owCDM_mnu}; see Sec. \ref{sec:mnu_double}).}
\label{table:result_owCDM_mnu_covar}
  \end{table*}

\begin{table*}
 \begin{center}
  \begin{tabular}{c c c c c c c c}
     \hline 
	&	$\Omega_m$	&			$H_0$	&			$\sigma_8$	&			$\Omega_k$	&			$w$ or $w_0$	&			$w_a$	&			$\Sigma{m_\mu}$(eV)									\\	\hline
$\Lambda$CDM	& $	0.310	\pm	0.010	$&$	67.6	\pm	0.8	$&$	0.828	\pm	0.019	$&$	0			$&$	-1			$&$	0			$&$	<				0.10	$ $(<	0.22	)		$\\	
o$\Lambda$CDM	& $	0.310	\pm	0.011	$&$	67.8	\pm	1.0	$&$	0.828	\pm	0.020	$&$	0.002	\pm	0.003	$&$	-1			$&$	0			$&$	<				0.13	$ $(<	0.27	)		$\\	
$w$CDM	& $	0.296	\pm	0.016	$&$	69.6	\pm	1.9	$&$	0.824	\pm	0.027	$&$	0			$&$	-1.11	\pm	0.10	$&$	0			$&$	<				0.26	$ $(<	0.52	)		$\\	
o$w$CDM	& $	0.297	\pm	0.017	$&$	69.8	\pm	2.2	$&$	0.816	\pm	0.033	$&$	0.001	\pm	0.004	$&$	-1.13	\pm	0.12	$&$	0			$&$	<				0.35	$ $(<	0.75	)		$\\	
$w_0w_a$CDM	& $	0.312	\pm	0.024	$&$	68.1	\pm	2.6	$&$	0.812	\pm	0.030	$&$	0			$&$	-0.88	\pm	0.24	$&$	-0.89	\pm	0.75	$&$	<				0.32	$ $(<	0.60	)		$\\	
o$w_0w_a$CDM	& $	0.310	\pm	0.026	$&$	68.3	\pm	3.3	$&$	0.809	\pm	0.034	$&$	-0.001	\pm	0.004	$&$	-0.91	\pm	0.29	$&$	-0.83	\pm	0.87	$&$	<				0.31	$ $(<	0.78	)		$\\	
\hline
\end{tabular}
 \end{center}
\caption{Constraints on cosmological parameters obtained by using the double-probe measurements presented in Table \ref{table:result_owCDM_mnu} and \ref{table:result_owCDM_mnu_covar} assuming dark energy models. We show 68\% 1-D marginalized constraints for all the parameters. We provide also 95\% constraints for the neutrino masses in the parentheses.
The units of $H_0$ and $\Sigma m_\nu$ are $\Hunit$ and eV respectively (see Sec. \ref{sec:mnu_double} and Fig. \ref{fig:mnu_doubleprobe}).
}
\label{table:assume_DEmodel_mnu_vary}
  \end{table*}

\begin{table*}
 \begin{center}
  \begin{tabular}{c c c c c c c c}
     \hline 
	&	$\Omega_m$	&			$H_0$	&			$\sigma_8$	&			$\Omega_k$	&			$w$ or $w_0$	&			$w_a$	&			$\Sigma{m_\mu}$(eV)								\\	\hline
$\Lambda$CDM	& $	0.306	\pm	0.009	$&$	68.0	\pm	0.7	$&$	0.803	\pm	0.017	$&$	0			$&$	-1			$&$	0			$&$	<				0.12	$ $(<	0.24	)	$\\	
o$\Lambda$CDM	& $	0.307	\pm	0.010	$&$	68.2	\pm	0.9	$&$	0.796	\pm	0.021	$&$	0.003	\pm	0.003	$&$	-1			$&$	0			$&$	<				0.19	$ $(<	0.37	)	$\\	
$w$CDM	& $	0.295	\pm	0.014	$&$	69.5	\pm	1.8	$&$	0.798	\pm	0.023	$&$	0			$&$	-1.10	\pm	0.10	$&$	0			$&$	<				0.27	$ $(<	0.53	)	$\\	
o$w$CDM	& $	0.296	\pm	0.015	$&$	70.1	\pm	2.3	$&$	0.781	\pm	0.033	$&$	0.003	\pm	0.004	$&$	-1.13	\pm	0.14	$&$	0			$&$	<				0.45	$ $(<	0.91	)	$\\	
$w_0w_a$CDM	& $	0.307	\pm	0.020	$&$	68.5	\pm	2.3	$&$	0.782	\pm	0.028	$&$	0			$&$	-0.92	\pm	0.22	$&$	-0.77	\pm	0.73	$&$	<				0.39	$ $(<	0.63	)	$\\	
o$w_0w_a$CDM	& $	0.302	\pm	0.021	$&$	69.4	\pm	2.8	$&$	0.775	\pm	0.034	$&$	0.002	\pm	0.004	$&$	-1.01	\pm	0.28	$&$	-0.53	\pm	0.88	$&$	<				0.47	$ $(<	0.93	)	$\\	
\hline
\end{tabular}
 \end{center}
\caption{Constraints on cosmological parameters obtained by using our double-probe measurements obtained with fixed $\Sigma m_\nu$ assuming dark energy models. We show 68\% 1-D marginalized constraints for all the parameters. We provide also 95\% constraints for the neutrino masses in the parentheses.
The units of $H_0$ and $\Sigma m_\nu$ are $\Hunit$ and eV respectively.
One can see that the results are very similar to Table \ref{table:assume_DEmodel_mnu_vary}, which showing our double probe measurements are insensitive to the $\Sigma m_\nu$ assumption
}
\label{table:assume_DEmodel_mnu_fixed}
  \end{table*}

\subsection{measuring neutrino mass using full likelihood analysis}
\label{sec:mnu_full_run}
\begin{figure}
\centering
\includegraphics[width=1 \columnwidth,clip,angle=0]{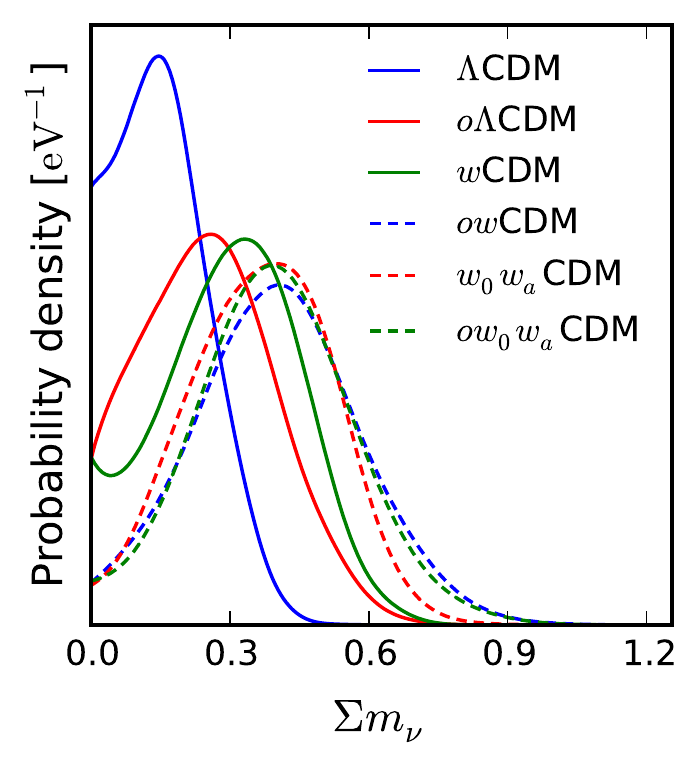}
\caption{
Probability density for $\Sigma m_\nu$
from the full-likelihood-analysis of the joint data set. $\Sigma m_\nu$ is one of the parameters to be constrained. Planck data including lensing with $A_{\rm L}=1$. The overall shape information of the monopole of the correlation function from the BOSS galaxy clustering is included (see Sec. \ref{sec:mnu_full_run} and Table \ref{table:fix_DEmodel_nopoly}).
}
\label{fig:mnu_3rd_cosmomc_nopoly}
\end{figure}

\begin{figure}
\centering
\includegraphics[width=1 \columnwidth,clip,angle=0]{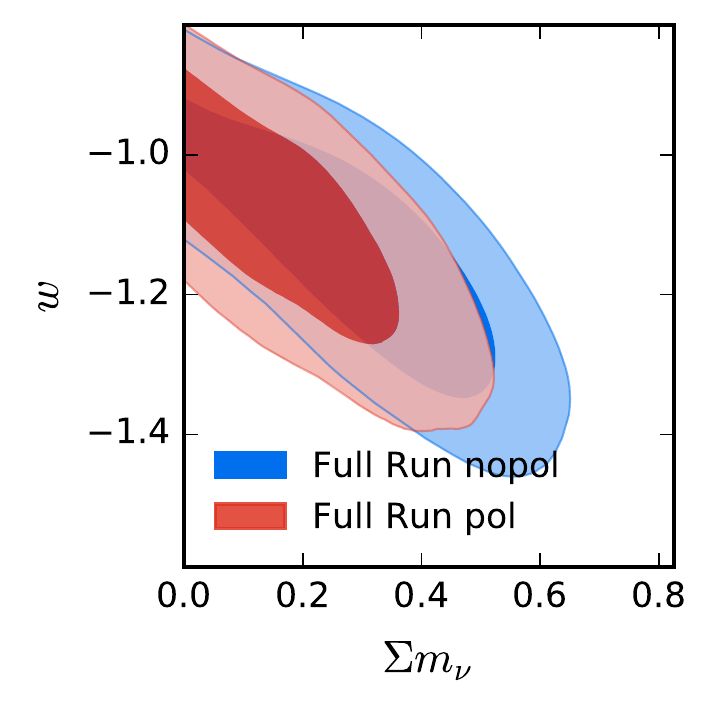}
\caption{
2D marginalized
  contours for $68\%$ and $95\%$ confidence level for $w$ and $\Sigma m_\nu$ ($w$CDM model assumed)
from Planck+BOSS. The blue contours are from full-likelihood-analysis without using a polynomial function to remove the overall shape information of monopole; the red contours are from the analysis removing overall shape information with a polynomial function. One can see that the overall shape information shift the $\Sigma m_\nu$ to a larger value.
}
\label{fig:w_mnu_compare_poly}
\end{figure}

\begin{figure}
\centering
\includegraphics[width=1 \columnwidth,clip,angle=0]{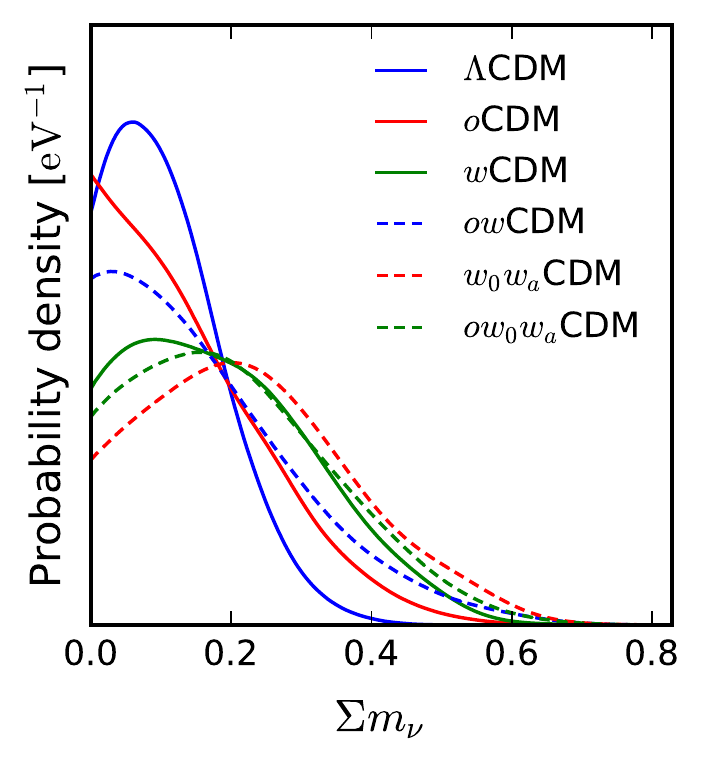}
\caption{
Probability density for $\Sigma m_\nu$
from the full-likelihood-analysis of the joint data set. $\Sigma m_\nu$ is one of the parameters to be constrained. Planck data includes lensing with $A_{\rm L}=1$. The overall shape information of the monopole of the correlation function from the BOSS galaxy clustering is removed with a polynomial function
 (see Sec. \ref{sec:mnu_full_run} and Table \ref{table:fix_DEmodel_poly_fixingAL}).
}
\label{fig:mnu_3rd_poly_fixingAL}
\end{figure}

We perform the same full MCMC analysis using the joint-full-likelihood of Planck and BOSS data as described in Sec. \ref{sec:full_run} 
to obtain the cosmological parameter constraints including $\Sigma m_\nu$. Table \ref{table:fix_DEmodel_nopoly} presents the results.
We show also the probability density for $\Sigma m_\nu$ in Fig. \ref{fig:mnu_3rd_cosmomc_nopoly}. We find more than 2 $\sigma$ detection 
of non zero $\Sigma m_\nu$ assuming the models without fixing $w$ to be -1. However, we find that the detection actually mainly comes from
the monopole of galaxy correlation function which is sensitive to some observational systematics, e.g. see \cite{Ross:2012qm,Chuang:2013wga}. 
Fig. \ref{fig:w_mnu_compare_poly} shows that the $\Sigma m_\nu$ detection decreases when adding a polynomial to remove the full shape information 
of monopole.
To be conservative, we run again the full MCMC analysis to obtain the constraint on $\Sigma m_\nu$ without including the full shape 
information and the results are presented in Table \ref{table:fix_DEmodel_poly_fixingAL}. 
The probability density for $\Sigma m_\nu$ is shown in Fig. \ref{fig:mnu_3rd_poly_fixingAL}. 
One can see the detections of $\Sigma m_\nu$ decrease. 
In addition, the upper limits in Fig. \ref{fig:mnu_3rd_poly_fixingAL} are lower than Fig. \ref{fig:mnu_doubleprobe} which are expected. 
Since we do not include the parameter $\Sigma m_\nu$ when summarising the information of double probe, the $\Sigma m_\nu$ constraint from $Planck$ is lost.

\begin{figure}
\centering
\includegraphics[width=1 \columnwidth,clip,angle=0]{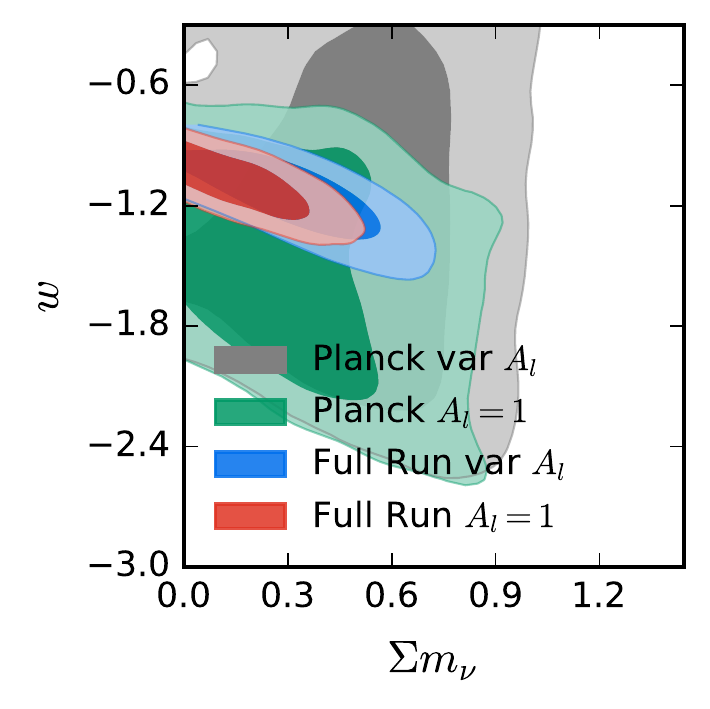}
\caption{
2D marginalized
  contours for $68\%$ and $95\%$ confidence level for $w$ and $\Sigma m_\nu$ ($w$CDM model assumed)
from full run methodology and Planck only for different lensing information used. Gray contours and green contours are from Planck only with varying $A_{\rm L}$ and fixing $A_{\rm L}=1$ respectively; the blue contours and the red contours are from Planck+BOSS with varying $A_{\rm L}$ and fixing $A_{\rm L}=1$ respectively using full-likelihood-analysis. One can see that $\Sigma m_\nu$ shift to a large value when varying $A_{\rm L}$ for both data combinations.
}
\label{fig:w_mnu_Alens}
\end{figure}

Table \ref{table:fix_DEmodel_poly_varyingAL} displays the constraints measured when allowing the CMB lensing amplitude parameter $A_{\rm L}$ to vary. Fig.
\ref{fig:w_mnu_Alens} shows the $Planck$ data shifts $\Sigma m_\nu$ measurement to higher values allowing 
a higher detection from the combined data analysis when allowing $A_{\rm L}$ free. Thus, we find again $ \sim$ 2 $\sigma$ detection even without accounting for the full shape of the monopole from the correlation function.

\begin{table*}
 \begin{center}
  \begin{tabular}{c c c c c c c c}
     \hline 
	&	$\Omega_m$	&			$H_0$	&			$\sigma_8$	&			$\Omega_k$	&			$w$ or $w_0$	&			$w_a$	&			$\Sigma{m_\mu}$(eV)									\\	\hline
$\Lambda$CDM	& $	0.308	\pm	0.011	$&$	67.7	\pm	0.9	$&$	0.801	\pm	0.017	$&$	0			$&$	-1			$&$	0			$&$	<				0.22	$ $(<	0.32	)		$\\	
o$\Lambda$CDM	& $	0.313	\pm	0.013	$&$	67.9	\pm	1.1	$&$	0.792	\pm	0.020	$&$	0.004	\pm	0.004	$&$	-1			$&$	0			$&$	0.25	_{	-0.17	}^{+	0.13	}$ $(<	0.49	)		$\\	
$w$CDM	& $	0.293	\pm	0.016	$&$	70.1	\pm	2.0	$&$	0.808	\pm	0.019	$&$	0			$&$	-1.15	\pm	0.11	$&$	0			$&$	0.30	_{	-0.14	}^{+	0.17	}$ $(<	0.52	)		$\\	
o$w$CDM	& $	0.299	\pm	0.019	$&$	70.0	\pm	2.4	$&$	0.795	\pm	0.021	$&$	0.004	\pm	0.004	$&$	-1.14	\pm	0.13	$&$	0			$&$	0.40	_{	-0.17	}^{+	0.17	}$ $\left(_{	-0.33	}^{+	0.34	}\right)$\\	
$w_0w_a$CDM	& $	0.316	\pm	0.023	$&$	67.8	\pm	2.5	$&$	0.785	\pm	0.023	$&$	0			$&$	-0.87	\pm	0.23	$&$	-0.96	\pm	0.68	$&$	0.36	_{	-0.15	}^{+	0.17	}$ $\left(_{	-0.29	}^{+	0.26	}\right)$\\	
o$w_0w_a$CDM	& $	0.313	\pm	0.026	$&$	68.4	\pm	2.8	$&$	0.787	\pm	0.027	$&$	0.002	\pm	0.004	$&$	-0.91	\pm	0.26	$&$	-0.82	\pm	0.77	$&$	0.39	_{	-0.15	}^{+	0.15	}$ $\left(_{	-0.32	}^{+	0.32	}\right)$\\		
\hline
\end{tabular}
 \end{center}
\caption{Constraints on cosmological parameters from the full-likelihood-analysis of the joint data set. $\Sigma m_\nu$ is one of the parameters to be constrained. Planck data includes lensing with $A_{\rm L}=1$. The overall shape information of the monopole of the correlation function from the BOSS galaxy clustering is included. We show 68\% 1-D marginalized constraints for all the parameters. We provide also 95\% constraints for the neutrino masses in the parentheses.
The units of $H_0$ and $\Sigma m_\nu$ are $\Hunit$ and eV respectively
 (see Sec. \ref{sec:mnu_full_run} and Fig. \ref{fig:mnu_3rd_cosmomc_nopoly}).
}
\label{table:fix_DEmodel_nopoly}
  \end{table*}

\begin{table*}
 \begin{center}
  \begin{tabular}{c c c c c c c c}
     \hline 
	&	$\Omega_m$	&			$H_0$	&			$\sigma_8$	&			$\Omega_k$	&			$w$ or $w_0$	&			$w_a$	&			$\Sigma{m_\mu}$(eV)									\\	\hline
$\Lambda$CDM	& $	0.309	\pm	0.011	$&$	67.7	\pm	0.9	$&$	0.808	\pm	0.015	$&$	0			$&$	-1			$&$	0			$&$	<				0.14	$ $(<	0.26	)		$\\	
o$\Lambda$CDM	& $	0.310	\pm	0.012	$&$	67.9	\pm	1.0	$&$	0.805	\pm	0.017	$&$	0.002	\pm	0.003	$&$	-1			$&$	0			$&$	<				0.18	$ $(<	0.36	)		$\\	
$w$CDM	& $	0.296	\pm	0.017	$&$	69.6	\pm	2.1	$&$	0.818	\pm	0.021	$&$	0			$&$	-1.11	\pm	0.11	$&$	0			$&$	<				0.25	$ $(<	0.42	)		$\\	
o$w$CDM	& $	0.300	\pm	0.019	$&$	69.1	\pm	2.2	$&$	0.813	\pm	0.021	$&$	0.001	\pm	0.004	$&$	-1.08	\pm	0.12	$&$	0			$&$	<				0.21	$ $(<	0.43	)		$\\	
$w_0w_a$CDM	& $	0.312	\pm	0.027	$&$	68.2	\pm	3.1	$&$	0.803	\pm	0.028	$&$	0			$&$	-0.91	\pm	0.27	$&$	-0.70	\pm	0.79	$&$	<				0.33	$ $(<	0.49	)		$\\	
o$w_0w_a$CDM	& $	0.311	\pm	0.025	$&$	68.0	\pm	2.7	$&$	0.803	\pm	0.026	$&$	0.000	\pm	0.004	$&$	-0.92	\pm	0.25	$&$	-0.59	\pm	0.78	$&$	<				0.28	$ $(<	0.45	)		$\\		
\hline
\end{tabular}
 \end{center}
\caption{Constraints on cosmological parameters from the full-likelihood-analysis of the joint data set. $\Sigma m_\nu$ is one of the parameters to be constrained. Planck data includes lensing with $A_{\rm L}=1$. The overall shape information of the monopole of the correlation function from the BOSS galaxy clustering is removed with a polynomial function. We show 68\% 1-D marginalized constraints for all the parameters. We provide also 95\% constraints for the neutrino masses in the parentheses.
The units of $H_0$ and $\Sigma m_\nu$ are $\Hunit$ and eV respectively
 (see Sec. \ref{sec:mnu_full_run} and Fig. \ref{fig:mnu_3rd_poly_fixingAL}).
}
\label{table:fix_DEmodel_poly_fixingAL}
  \end{table*}

 \begin{table*}
 \begin{center}
  \begin{tabular}{c c c c c c c c c}
     \hline 
	&	$\Omega_m$	&			$H_0$	&			$\sigma_8$	&			$\Omega_k$	&			$w$ or $w_0$	&			$w_a$	&			$\Sigma{m_\mu}$(eV)		&								$A_L$			\\	\hline
$\Lambda$CDM	& $	0.308	\pm	0.011	$&$	67.7	\pm	0.9	$&$	0.782	\pm	0.026	$&$	0			$&$	-1			$&$	0			$&$	0.17	_{	-0.13	}^{+	0.08	}$ $(<	0.34	)		$&$	1.07	\pm	0.06	$\\	
o$\Lambda$CDM	& $	0.314	\pm	0.013	$&$	67.9	\pm	1.0	$&$	0.752	\pm	0.037	$&$	0.005	\pm	0.004	$&$	-1			$&$	0			$&$	0.34	_{	-0.22	}^{+	0.17	}$ $(<	0.66	)		$&$	1.12	\pm	0.07	$\\	
$w$CDM	& $	0.290	\pm	0.019	$&$	70.4	\pm	2.5	$&$	0.781	\pm	0.032	$&$	0			$&$	-1.16	\pm	0.14	$&$	0			$&$	0.33	_{	-0.18	}^{+	0.16	}$ $(<	0.60	)		$&$	1.10	\pm	0.07	$\\	
o$w$CDM	& $	0.300	\pm	0.023	$&$	69.8	\pm	2.8	$&$	0.754	\pm	0.041	$&$	0.005	\pm	0.005	$&$	-1.11	\pm	0.15	$&$	0			$&$	0.44	_{	-0.22	}^{+	0.23	}$ $(<	0.81	)		$&$	1.13	\pm	0.07	$\\	
$w_0w_a$CDM	& $	0.292	\pm	0.031	$&$	70.4	\pm	3.9	$&$	0.781	\pm	0.037	$&$	0			$&$	-1.15	\pm	0.34	$&$	-0.09	\pm	0.94	$&$	0.32	_{	-0.20	}^{+	0.18	}$ $(<	0.61	)		$&$	1.10	\pm	0.06	$\\	
o$w_0w_a$CDM	& $	0.292	\pm	0.030	$&$	70.8	\pm	3.7	$&$	0.763	\pm	0.044	$&$	0.004	\pm	0.005	$&$	-1.18	\pm	0.32	$&$	0.11	\pm	0.94	$&$	0.42	_{	-0.22	}^{+	0.22	}$ $(<	0.77	)		$&$	1.14	\pm	0.09	$\\	
\hline
\end{tabular}
 \end{center}
\caption{Constraints on cosmological parameters from the full-likelihood-analysis from the joint data set. Both $\Sigma m_\nu$ and $A_{\rm L}$ are the parameters to be constrained. The overall shape information of the monopole of the correlation function from the BOSS galaxy clustering is removed with a polynomial function. We show 68\% 1-D marginalized constraints for all the parameters. We provide also 95\% constraints for the neutrino masses in the parentheses.
The units of $H_0$ and $\Sigma m_\nu$ are $\Hunit$ and eV respectively
 (see Sec. \ref{sec:mnu_full_run} and Fig. \ref{fig:mnu_3rd_poly_varyAL}).
}
\label{table:fix_DEmodel_poly_varyingAL}
  \end{table*}

\begin{figure}
\centering
\includegraphics[width=1 \columnwidth,clip,angle=0]{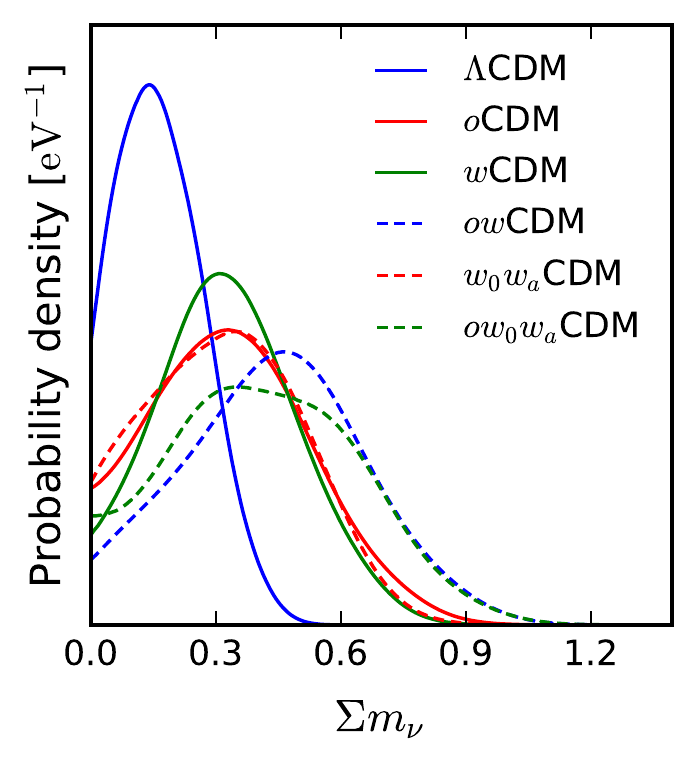}
\caption{
Probability density for $\Sigma m_\nu$
from full-likelihood-analysis from the joint data set. Both $\Sigma m_\nu$ and $A_{\rm L}$ are the parameters to be constrained. The overall shape information of the monopole of the correlation function from the BOSS galaxy clustering is removed with a polynomial function (see Sec. \ref{sec:mnu_full_run} and Table \ref{table:fix_DEmodel_poly_varyingAL}). One can see that the maximum of $\Sigma m_\nu$ increases comparing to the cases with fixing $A_{\rm L}=1$ (see Fig. \ref{fig:mnu_3rd_poly_fixingAL}).
}
\label{fig:mnu_3rd_poly_varyAL}
\end{figure}

\subsection{measure neutrino mass using measurements from single probe analysis}
\label{sec:mnu_single}

\begin{figure}
\centering
\includegraphics[width=1 \columnwidth,clip,angle=0]{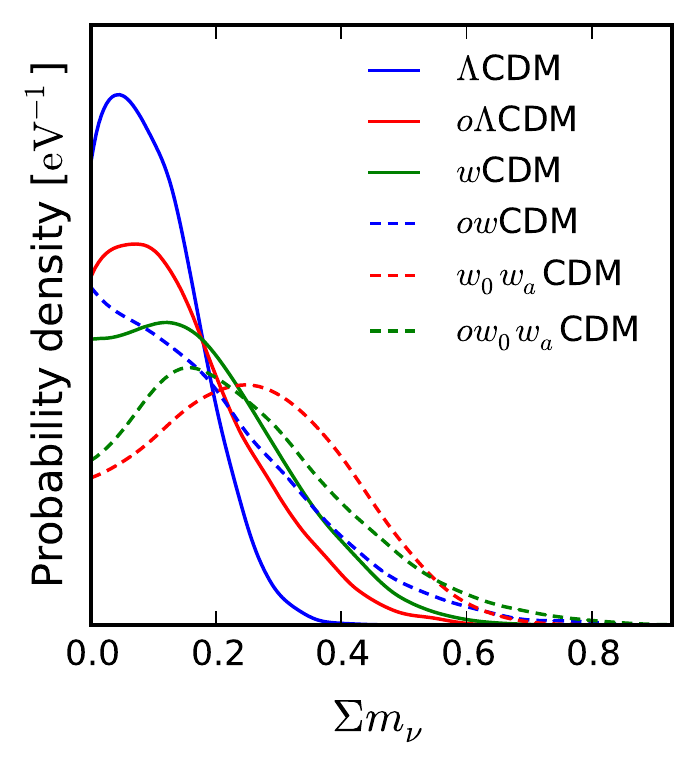}
\caption{
Probability density for $\Sigma m_\nu$
from the single probe measurements provided by Chuang et al. 2016 (companion paper) combining with Planck data (with fixing $A_{\rm L}=1$). All the measurements are consistent with $\Sigma m_\nu=0$
 (see Sec. \ref{sec:mnu_full_run} and Table \ref{table:DEmodels_mnu_singleprobe}). 
}
\label{fig:mnu_singleprobe}
\end{figure}

We use the single probe measurement provided by Chuang et al. (companion paper) combining with Planck (fixing $A_{\rm L}=1$) and obtain the constraint of $\Sigma m_\nu$. 
Table \ref{table:DEmodels_mnu_singleprobe} shows the cosmological parameter constraints including $\Sigma m_\nu$ for different dark 
energy models. The probability densities for $\Sigma m_\nu$ are shown in Fig. \ref{fig:mnu_singleprobe}. One can see that it is 
consistent with Fig. \ref{fig:mnu_3rd_poly_fixingAL}. We have checked that there would be some detection of neutrino mass while allowing $A_{\rm L}$ to be free as seen in the case of full-likelihood analysis (see Sec. \ref{sec:mnu_full_run}).

\begin{figure}
\centering
\includegraphics[width=1 \columnwidth,clip,angle=0]{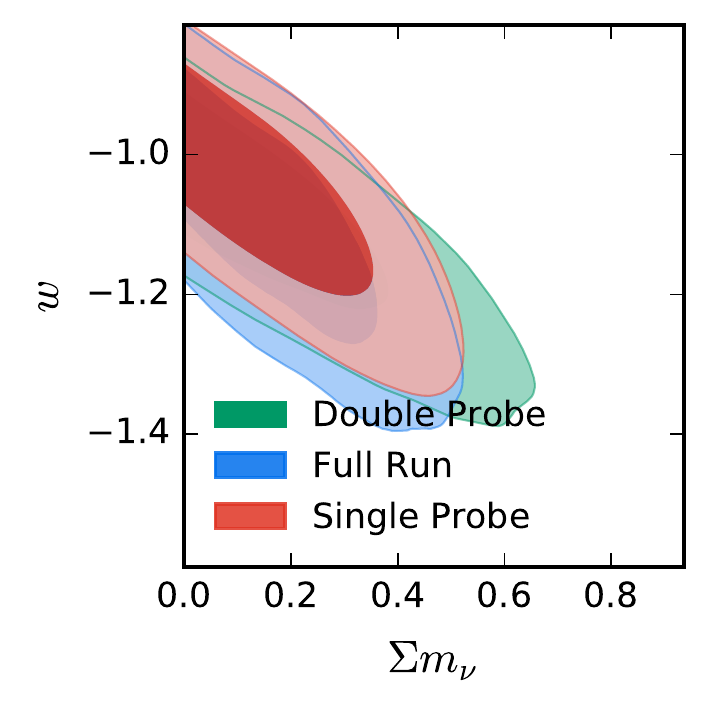}
\caption{
Comparison of 2D contours for $68\%$ and $95\%$ confidence level on $\Sigma m_\nu$ and $w$
from the double probe, single probe, and full-likelihood-analysis approaches. 
One can see that the constraint on $\Sigma m_\nu$ from the double probe approach is weaker which is expected. The difference comes from the fact that we do not include $\Sigma m_\nu$ into our summarized set of parameters, so information from $Planck$ is lost.
}
\label{fig:mnu_single_double_full_probe}
\end{figure}

Fig. \ref{fig:mnu_single_double_full_probe} presents the comparison between the three different methodologies.
The three approaches agree very well with some subtle differences. 
One can see that the constraint on $\Sigma m_\nu$ from the double probe approach is weaker which is expected. The difference comes from the fact that we do not include $\Sigma m_\nu$ into our summarized set of parameters, so that information from $Planck$ is lost. On the other hand, both single probe and full-likelihood analysis include full $Planck$ 
information and their measurements are very similar.

\begin{table*}
\begin{center}
\begin{tabular}{lrrrrrrr}
\hline
	&	$\Omega_m$	&			$H_0$	&			$\sigma_8$	&			$\Omega_k$	&			$w$ or $w_0$	&			$w_a$	&			$\Sigma{m_\mu}$(eV)									\\	\hline
$\Lambda$CDM	& $	0.310	\pm	0.010	$&$	67.6	\pm	0.8	$&$	0.809	\pm	0.014	$&$	0			$&$	-1			$&$	0			$&$	<				0.14	$ $(<	0.24	)		$\\	
o$\Lambda$CDM	& $	0.313	\pm	0.011	$&$	67.6	\pm	0.9	$&$	0.804	\pm	0.016	$&$	0.002	\pm	0.004	$&$	-1			$&$	0			$&$	<				0.19	$ $(<	0.37	)		$\\	
$w$CDM	& $	0.303	\pm	0.014	$&$	68.7	\pm	1.7	$&$	0.812	\pm	0.017	$&$	0			$&$	-1.08	\pm	0.09	$&$	0			$&$	<				0.24	$ $(<	0.42	)		$\\	
o$w$CDM	& $	0.305	\pm	0.014	$&$	68.6	\pm	1.6	$&$	0.809	\pm	0.018	$&$	0.001	\pm	0.004	$&$	-1.06	\pm	0.10	$&$	0			$&$	<				0.25	$ $(<	0.48	)		$\\	
$w_0w_a$CDM	& $	0.314	\pm	0.021	$&$	67.8	\pm	2.2	$&$	0.800	\pm	0.022	$&$	0			$&$	-0.91	\pm	0.22	$&$	-0.70	\pm	0.75	$&$	0.26	_{	-0.18	}^{+	0.13	}$ $(<	0.51	)		$\\	
o$w_0w_a$CDM	& $	0.315	\pm	0.020	$&$	67.6	\pm	2.1	$&$	0.799	\pm	0.022	$&$	-0.001	\pm	0.004	$&$	-0.89	\pm	0.21	$&$	-0.77	\pm	0.74	$&$	0.24	_{	-0.22	}^{+	0.08	}$ $(<	0.55	)		$\\	
\hline
\end{tabular}
\end{center}
\caption{ 
The cosmological constraints including total mass of neutrinos from the single probe measurements provided by Chuang et al. 2016 (companion paper) combining with Planck data assuming different dark energy models.
We show 68\% 1-D marginalized constraints for all the parameters. We provide also 95\% constraints for the neutrino masses in the parentheses.
The units of $H_0$ and $\Sigma m_\nu$ are $\Hunit$ and eV respectively
 (see Sec. \ref{sec:mnu_full_run} and Fig.\ref{fig:mnu_singleprobe}).
} \label{table:DEmodels_mnu_singleprobe}
\end{table*}

\subsection{combination with supernovae type Ia data}
\label{sec:mnu_JLA_full_run}

\begin{figure}
\centering
\includegraphics[width=1 \columnwidth,clip,angle=0]{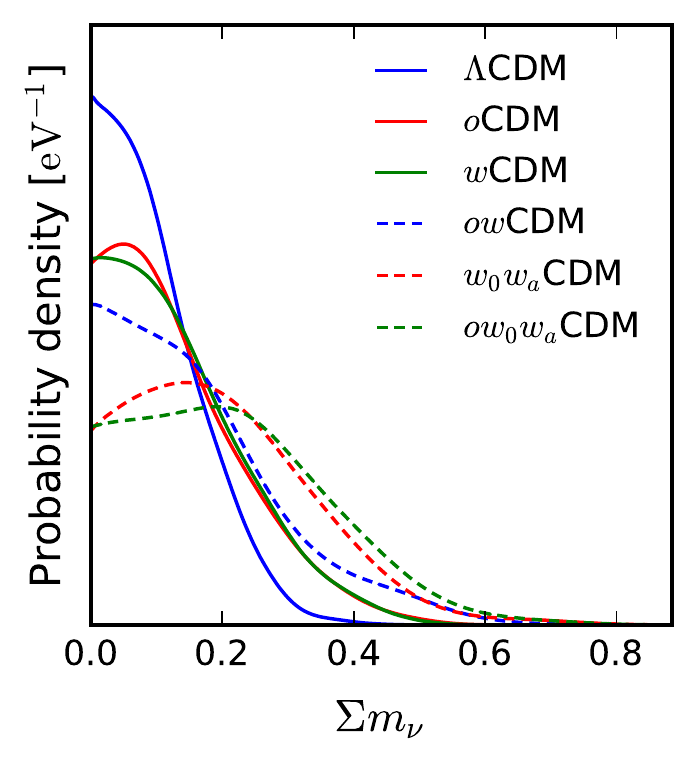}
\caption{
Probability density for $\Sigma m_\nu$
from the full likelihood analysis measurement for joint and JLA data sets. We assume lensing likelihood with fixed $A_{\rm L}=1$. 
All the measurements are consistent with $\Sigma m_\nu=0$
 (see Sec. \ref{sec:mnu_full_run} and Table \ref{table:fix_DEmodel_poly_fixingAL_JLA}). 
}
\label{fig:mnu_full_run_JLA}
\end{figure}

\begin{figure}
\centering
\includegraphics[width=1 \columnwidth,clip,angle=0]{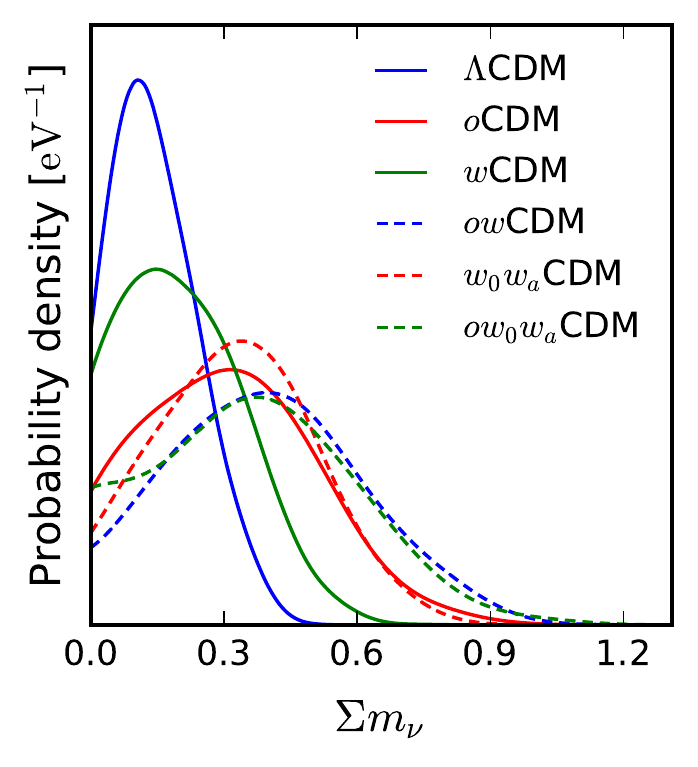}
\caption{
Probability density for $\Sigma m_\nu$
from the full likelihood analysis measurement for joint and JLA data sets. We assume lensing likelihood with variable $A_{\rm L}$
 (see Sec. \ref{sec:mnu_full_run} and Table \ref{table:fix_DEmodel_poly_varyingAL_JLA}).  
}
\label{fig:mnu_full_run_JLA_Alens}
\end{figure}

We combine our measurements using the full likelihood approach with those from supernovae Ia (SNIa) data, Joint Light-curve 
Analysis (JLA) \citep{Betoule:2014frx}. As seen in Fig. \ref{fig:w_mnu_SN}, SN data breaks some degeneracies providing tighter
constraints on $\Sigma m_\nu$. Results can be found in table \ref{table:fix_DEmodel_poly_fixingAL_JLA} and Fig. \ref{fig:mnu_full_run_JLA}) 
for the
case of fixing $A_{\rm L}=1$ and 
table \ref{table:fix_DEmodel_poly_varyingAL_JLA} and Fig. \ref{fig:mnu_full_run_JLA_Alens}) for the case of varying $A_{\rm L}$. When adding SN1a data, we get tighter upper limits, e.g.
$\Sigma m_\nu<0.12$ against $\Sigma m_\nu<0.14$ in $\Lambda CDM$ with $A_{\rm L}=1$. 
We point out that the constraints we obtained are still not sufficient to distinguish between normal and inverted hierarchy.


\begin{table*}
 \begin{center}
  \begin{tabular}{c c c c c c c c}
     \hline 
	&	$\Omega_m$	&			$H_0$	&			$\sigma_8$	&			$\Omega_k$	&			$w$ or $w_0$	&			$w_a$	&			$\Sigma{m_\mu}$(eV)									\\	\hline
$\Lambda$CDM	& $	0.309	\pm	0.010	$&$	67.7	\pm	0.8	$&$	0.810	\pm	0.014	$&$	0			$&$	-1			$&$	0			$&$	<				0.12	$ $(<	0.24	)		$\\	
o$\Lambda$CDM	& $	0.309	\pm	0.010	$&$	67.9	\pm	0.9	$&$	0.807	\pm	0.016	$&$	0.001	\pm	0.004	$&$	-1			$&$	0			$&$	<				0.17	$ $(<	0.33	)		$\\	
$w$CDM	& $	0.305	\pm	0.012	$&$	68.2	\pm	1.2	$&$	0.812	\pm	0.016	$&$	0			$&$	-1.04	\pm	0.05	$&$	0			$&$	<				0.17	$ $(<	0.33	)		$\\	
o$w$CDM	& $	0.307	\pm	0.013	$&$	68.3	\pm	1.4	$&$	0.808	\pm	0.019	$&$	0.001	\pm	0.004	$&$	-1.03	\pm	0.06	$&$	0			$&$	<				0.20	$ $(<	0.43	)		$\\	
$w_0w_a$CDM	& $	0.309	\pm	0.014	$&$	68.2	\pm	1.3	$&$	0.807	\pm	0.019	$&$	0			$&$	-0.92	\pm	0.12	$&$	-0.64	\pm	0.56	$&$	<				0.26	$ $(<	0.43	)		$\\	
o$w_0w_a$CDM	& $	0.310	\pm	0.013	$&$	68.0	\pm	1.3	$&$	0.803	\pm	0.019	$&$	0.000	\pm	0.004	$&$	-0.91	\pm	0.11	$&$	-0.63	\pm	0.59	$&$	<				0.27	$ $(<	0.46	)		$\\	
\hline
\end{tabular}
 \end{center}
\caption{Constraints on cosmological parameters from the full-likelihood-analysis of the joint (Planck and BOSS dr12) and JLA data sets assuming variable $\Sigma m_\nu$. Planck data includes lensing with $A_{\rm L}=1$. The overall shape information of the monopole of the correlation function from the BOSS galaxy clustering is removed with a polynomial function.
We show 68\% 1-D marginalized constraints for all the parameters. We provide also 95\% constraints for the neutrino masses in the parentheses.
The units of $H_0$ and $\Sigma m_\nu$ are $\Hunit$ and eV respectively
 (see Sec. \ref{sec:mnu_full_run} and Fig. \ref{fig:mnu_full_run_JLA}). 
}
\label{table:fix_DEmodel_poly_fixingAL_JLA}
  \end{table*}

 \begin{table*}
 \begin{center}
  \begin{tabular}{c c c c c c c c c}
     \hline 
	&	$\Omega_m$	&			$H_0$	&			$\sigma_8$	&			$\Omega_k$	&			$w$ or $w_0$	&			$w_a$	&			$\Sigma{m_\mu}$(eV)		&								$A_L$			\\	\hline
$\Lambda$CDM	& $	0.307	\pm	0.010	$&$	67.8	\pm	0.8	$&$	0.784	\pm	0.026	$&$	0			$&$	-1			$&$	0			$&$	0.15	_{	-0.13	}^{+	0.06	}$ $(<	0.32	)		$&$	1.07	\pm	0.06	$\\	
o$\Lambda$CDM	& $	0.311	\pm	0.013	$&$	68.0	\pm	1.1	$&$	0.755	\pm	0.037	$&$	0.005	\pm	0.005	$&$	-1			$&$	0			$&$	0.32	_{	-0.23	}^{+	0.16	}$ $(<	0.63	)		$&$	1.12	\pm	0.08	$\\	
$w$CDM	& $	0.306	\pm	0.012	$&$	68.2	\pm	1.2	$&$	0.779	\pm	0.030	$&$	0			$&$	-1.04	\pm	0.06	$&$	0			$&$	0.21	_{	-0.18	}^{+	0.09	}$ $(<	0.44	)		$&$	1.08	\pm	0.07	$\\	
o$w$CDM	& $	0.310	\pm	0.012	$&$	68.5	\pm	1.3	$&$	0.748	\pm	0.038	$&$	0.006	\pm	0.004	$&$	-1.04	\pm	0.06	$&$	0			$&$	0.40	_{	-0.25	}^{+	0.19	}$ $(<	0.76	)		$&$	1.13	\pm	0.08	$\\	
$w_0w_a$CDM	& $	0.310	\pm	0.013	$&$	68.1	\pm	1.2	$&$	0.769	\pm	0.035	$&$	0			$&$	-0.93	\pm	0.12	$&$	-0.70	\pm	0.61	$&$	0.33	_{	-0.18	}^{+	0.16	}$ $(<	0.61	)		$&$	1.09	\pm	0.07	$\\	
o$w_0w_a$CDM	& $	0.310	\pm	0.016	$&$	68.5	\pm	1.6	$&$	0.756	\pm	0.037	$&$	0.004	\pm	0.005	$&$	-0.97	\pm	0.14	$&$	-0.41	\pm	0.67	$&$	0.38	_{	-0.27	}^{+	0.20	}$ $(<	0.74	)		$&$	1.12	\pm	0.08	$\\	
\hline
\end{tabular}
 \end{center}
\caption{Constraints on cosmological parameters from the full-likelihood-analysis of the joint (Planck and BOSS dr12) and JLA data sets assuming variable $\Sigma m_\nu$. Planck data includes lensing varying $A_{\rm L}$. The overall shape information of the monopole of the correlation function from the BOSS galaxy clustering is removed with a polynomial function.
We show 68\% 1-D marginalized constraints for all the parameters. We provide also 95\% constraints for the neutrino masses in the parentheses.
The units of $H_0$ and $\Sigma m_\nu$ are $\Hunit$ and eV respectively
 (see Sec. \ref{sec:mnu_full_run} and Fig. \ref{fig:mnu_full_run_JLA_Alens}). 
}
\label{table:fix_DEmodel_poly_varyingAL_JLA}
  \end{table*}

